\documentclass[showpacs,preprintnumbers,amsmath,amssymb,superscriptaddress]{revtex4}

\usepackage{mathrsfs,amsmath}
\usepackage[scr=boondox]{mathalfa}
\usepackage{comment}
\usepackage{graphicx, float}
\usepackage{dcolumn}
\usepackage{bm}
\usepackage{epstopdf}
\usepackage{doi,hyperref,url}
\usepackage[usenames, dvipsnames]{color}

\usepackage[caption=false]{subfig}

\begin{document}

\title{Fast prediction and evaluation of eccentric inspirals using reduced-order models}
\author{D\'{a}niel Barta}
 \email{barta.daniel@wigner.mta.hu}
 \affiliation{%
    Institute for Particle and Nuclear Physics, Wigner Research Centre for Physics, Hungarian Academy of Sciences,
    Konkoly-Thege Mikl\'{o}s \'{u}t 29-33., H-1121 Budapest, Hungary
  }%
\author{M\'{a}ty\'{a}s Vas\'{u}th}
\email{vasuth.matyas@wigner.mta.hu}
\affiliation{%
	Institute for Particle and Nuclear Physics, Wigner Research Centre for Physics, Hungarian Academy of Sciences,
	Konkoly-Thege Mikl\'{o}s \'{u}t 29-33., H-1121 Budapest, Hungary
}

\date{\today}

\begin{abstract}
A large number of theoretically predicted waveforms are required by matched-filtering searches for the gravitational-wave signals produced by compact binary coalescence. In order to substantially alleviate the computational burden in gravitational-wave searches and parameter estimation without degrading the signal detectability, we propose a novel reduced-order-model (ROM) approach with applications to adiabatic 3PN-accurate inspiral waveforms of sources that evolve on either highly or slightly eccentric orbits. We provide a singular-value decomposition-based reduced-basis method in the frequency domain to generate reduced-order approximations of any gravitational waves with acceptable accuracy and precision within the parameter range of the model. We construct efficient reduced bases comprised of a relatively small number of the most relevant waveforms over 3-dimensional parameter-space covered by the template bank (total mass $2.15M_{\odot} \leq M \leq 215M_{\odot}$, mass ratio $0.01 \leq q \leq 1$, and initial orbital eccentricity $0 \leq e_{0} \leq 0.95$). The ROM is designed to predict signals in the frequency band from 10 Hz to 2 kHz for aLIGO and aVirgo design sensitivity. Beside moderating the data reduction, finer sampling of fiducial templates improves the accuracy of surrogates. Considerable increase in the speedup from several hundreds to thousands can be achieved by evaluating surrogates for low-mass systems especially when combined with high-eccentricity. 
\end{abstract}

\maketitle

\section{\label{sec:intro}Introduction}
This present work is a response to the growing demand for computationally-efficient generation of eccentric waveform families in gravitational-wave (GW) searches. To the extent of our knowledge, surrogate model building for this particular family of waveforms has not yet been tested. As ROM techniques have proved exceedingly efficient for other models (such as for aligned-spin BBHs), we thus anticipate similar benefits of extremely large speedups in the time-consuming process of generating eccentric waveform. Our aim is to demonstrate its exceptional expedience by design and to offer a novel and practical way to dramatically accelerate parameter estimations. \\

Compact binary coalescences (CBCs) in binary compact objects, such as stellar-mass binary black holes (BBHs) and/or binary neutron stars (BNSs), are among the most promising GWs sources for ground-based GW detectors. \cite{sathyaprakash} \nocite{sathyaprakash} Binaries that evolved through typical main sequence evolution \cite{kalogera} are expected to shed their formation eccentricities over time due to gravitational radiation reaction. For this reason, isolated compact binaries are commonly assumed to move on quasicircular orbits by the time they spiral into the sensitive frequency band of terrestrial GW observatories. \cite{peters,antonini} Some relatively young sources, nevertheless, which had too short time for the gravitational radiation reaction to completely circularize their orbits retain some residual eccentricity. \cite{shapiro} Therefore, CBC inspirals with non-negligible orbital eccentricities are plausible sources. \cite{peters} Some results \cite{pierro,martel} support the qualitative conclusion that neglecting residual orbital eccentricities (even small ones) in CBCs may seriously deteriorate matched-filter detection performance. A number of possible astrophysical scenarios and mechanisms allows the formation of observationally relevant eccentric ultracompact binaries (see in \cite{huerta,yunes,konigsdorffer}). Short-period CBCs may form by dynamical capture in dense stellar enviroments, present in both galactic central regions and globular clusters, or by tidal capture of compact object by NSs which is described in great detail in \cite{samsing,oleary,lee}. Stable hierarchical triple star-systems may form in globular clusters where multi-body interactions are involved. It has been estimated that $\sim 30\%$ of binaries formed in systems where the Kozai resonance increased the eccentricity of the inner binary will have initial eccentricities $e_{0} > 0.1$ when they enter the frequency window of the aLIGO. \cite{brown} Great majority ($\sim 90\%$) of stellar-mass BH binaries formed by scattering in galactic cores containing a supermassive BH have $e_{0} > 0.9$, where $e_{0}$ denotes the initial eccentricity of the binary by the time it enters the lower part of the frequency band of detectors. \cite{oleary} Roughly $0.1-10$ eccentric inspiral events per year up to redshift $z \sim 0.2$ are anticipated to be discovered by aLigo-type observatories. \cite{huerta} One of the key goals of GW observatories is to measure the intrinsic parameters of coalescing BNSs. Moreover, Favata (2014) pointed out that neglecting initial eccentricities $e_{0} \gtrsim 0.002$ causes systematic errors that exceed statistical errors in aLIGO measurements. \cite{favata} Since the phasing of the GW signal is significantly more important for parameter estimation, and eccentricity modify the phasing beginning at 1.5 and 0PN orders, eccentricity corrections to the SPA (stationary phase approximation) phase have to be included at leading order.

Putative frequency modulated GW signals (also known as `\emph{chirps}') from CBC inspirals will be buried in the noisy data streams of the advanced detectors. Data analysis of targeted search techniques operate by matched-filtering to extract any possible signal from the white Gaussian noise by cross-correlating the discrete-time sequences of the detector data against a large set of theoretical \emph{waveform templates} (or \emph{filters}) which approximate potential astrophysical signals. \cite{smith} The utility of this technique rests partly on how accurately the applied template waveforms model the signal being sought. Stellar-mass BBHs and BNSs in the inspiral regime are adequately described by high-order \emph{Post-Newtonian} (PN) waveform templates. PN approximations \cite{blanchet} are expansions of Einstein field equations to any specified order in a small parameter $v/c$ which provide a powerful formalism for modelling CBCs during the inspiral phase, when the orbital speed of the binary $v$ is much smaller than the speed of light $c$. \cite{brown} A PN extension  of order $(v/c)^{n}$ to the Newtonian expression of gravity is said to be of $(n/2)$PN order. We construct PN template families by making use of a fast and accurate computational tool, the \emph{CBwaves} software, developed by the Virgo Group at Wigner RCP. The 3PN-accurate equations of motion (and the spin precession equations if needed) of the orbiting bodies are integrated by a fourth-order Runge--Kutta method while the far-zone radiation field is determined by a simultaneous evaluation of analytic waveforms. The waveforms involve all high-order relativistic contributions for generic eccentric orbits up to 2PN-order accuracy. \cite{csizmadia}\nocite{CBwaves-web} The choice of PN template families used in this paper is also motivated by the very fact that such templates are available in the \emph{LSC Algorithms Library} (LAL) which applies some of them in targeted searches, although current searches do not exceed 2PN order. \cite{buonanno}

GWs are parametrized by a set of \emph{intrinsic} and \emph{extrinsic parameters} $\bm{\lambda} = \bm{\lambda}_{\text{extrinsic}} + \bm{\lambda}_{\text{intrinsic}}$, associated with the astrophysical model of their respective sources. The earlier are intrinsic to the source (such as the masses, spins and eccentricity of the compact objects) while the extrinsic parameters are those which depend on the relative location of the source with respect to the detector (such as the time of arrival $t_{0}$ of the signal at the detector and the phase of the signal $\phi_{0}$ at a reference time $t_{0}$). Each template has a specific set of values for its parameters which are hereinafter collectively referred to as \emph{model parameters}. A collection of points in a \emph{$p$-dimensional parameter space}, provided that $p$ is the number of model parameters, is called a \emph{template bank} (or \emph{template grid}). \cite{allen} A template bank generated with \emph{minimal match} $MM$ could contain a large number of templates that scales as $L \sim (1-MM)^{-p/2}$. The number of templates $L$ required for correlations grows rapidly with $p$ and the number of GW cycles $L_{\text{cyc}}$. \cite{owen} A fully coherent GW search for a CBC with $p = 8$ parameters lasting for $L_{\text{cyc}} = 10^{5}$ cycles would require as much as $L = 10^{40}$ waveform model evaluations. \cite{babak} 

Over the last three decades methods have been developed for setting up template banks which minimize the computational cost in GW searches without degrading the signal detectability, measured by the \emph{signal-to-noise ratio} (SNR). \cite{sathyaprakash2,sathyaprakash3,dhurandhar} Since the 1990s a method most feasible for small-dimensional parameter space ($p = 2,\, 3$, or $4$ at most) has been popular to address the problem of \emph{template placement} by associating the parameter space with a positive-definite metric space. In this geometric framework, the metric measures the fractional loss in squared SNR of a predicted signal (at one point in the parameter space) filtered through the optimal waveform template corresponding to a nearby point in the parameter space. \cite{owen2} In 2009 a template placement algorithm was developed that is suitable for any number of dimensions, provided that the metric distance between two points in the parameter space is large or well-defined. \cite{harry}

Beside the issue of ensembling sufficiently large template banks, \emph{parameter estimation} (PE) carries a number of challenges unique to large data sets. The exploration of the parameter space of BBHs relies on numerical relativity (NR) simulations of the field equations to discover how such mergers evolve. \cite{canizares} Even a very coarse survey of the parameter space would require an enormous number, typically $L = 10^{6}-10^{7}$ \cite{aasi}, of expensive NR simulations which impose a computationally insuperable obstacle. The required number is in fact subtantially greater than the combined number of all simulations ever performed by each and every NR group \cite{mroue,pekowsky}. Consequently, techniques which can estimate the astrophysical parameters fast and accurately are needed to overcome this computational bottleneck. \cite{field}

\emph{Reduced-order modeling} or \emph{model order reduction} is a practical mathematical tool to extract the fundamental features of a computationally demanding high-order model through exploiting only a reduced set of information. Investigations \cite{field2,field3,cannon,cannon2} over the last few years have revealed that GW templates exhibit significant redundancy in the parameter space, suggesting that the amount of information required to represent a fiducial waveform model is appreciably smaller than commonly	anticipated. The reduction of information content is achieved through expressing the essential information by means of only a remarkably few, reduced number of representative waveforms $r \ll L$ to construct a \emph{reduced-order model} (ROM) also known as a \emph{surrogate model}. ROMs provide compressed approximations of any selected waveforms within the same physical model. They are projection-based techniques that aim to lower the computational complexity in the simulations by mapping the original full-order model (FOM) onto an appropriate subspace of much lower dimension spanned by a \emph{reduced-order basis} (RB). To find these representative waveforms that constitutes the RB several methods, including \emph{singular value decompsition} (SVD) and \emph{greedy} methods have been proposed, usually combined with the \emph{empirical interpolation method} (EIM). \cite{field2,maday} SVD-based methods have been applied in Ref. \cite{cannon2,cannon3,cannon4} to interpolate time-domain inspiral waveforms. We are going to provide an effcient (fast and accurate) representations of approximated waveforms for any desired parameter values within the model by using the information provided by only $r$ RB waveforms instead of the total number $L$. \cite{cannon,privitera} The SVD-based approach to significantly accerelate PE process used in Ref. \cite{smith} is to directly interpolate the likelihood function over a significant portion of the parameter space. Moreover there is yet another method, presented in Ref. \cite{canizares,canizares2}, that defines special \emph{reduced-order quadrature} (ROQ) rules to assist in fast likelihood evaluation. \\

The rest of the paper is organized as follows: Sec. \ref{sec:fiducial-models} deals with the procedure for generating fiducial PN waveforms by CBwaves, with respect to the statistics of the cost of computing individual waveforms to estimate the total cost of building template banks. Sec. \ref{sec:grids} proposes the simplest strategy (regular spacing) for template placement in the intrinsic parameter space, followed by the representation of the fiducial waveform templates on a common, finely sampled and regularly spaced frequency grid. Sec. \ref{sec:rom} gives a general description of our approach to construct efficient ROM assembled from the reduced bases and of its characteristic features, particularly the truncation error. Sec. \ref{sec:efficiency} is dedicated to assess the overall performance of the ROM, including the accuracy of the surrogate model and its computational cost relative to that of the fiducial model. Conclusions, remarks, limitations and an outlook for future research will be given in Sec. \ref{sec:conclusion}.

\section{Fiducial waveform models}\label{sec:fiducial-models}
Current searches for GWs from NS and stellar-mass BH binaries use restricted stationary-phase approximations to the Fourier transform of 3.5PN-accurate circular inspiral-only waveforms, such as spin-aligned \emph{TaylorF2} or \emph{SpinTaylorT4}. \cite{brown} The first part of this section describes a procedure for constructing PN eccentric waveforms by \emph{CBwaves} model in the time domain. The second part deals with the statistics of the cost of computing individual time-domain (TD) waveforms, drawn from a relatively large number of sample points in a finite-sample distribution.

\subsection{Construction of eccentric post-Newtonian waveform templates}\label{sec:eccentric-waveform}
The \emph{CBwaves} open-source software was developed by the Virgo Group at Wigner Research Centre for Physics with the intent of providing efficient computational tool capable of generating gravitational waveforms produced by generic spinning binary configurations moving on eccentric closed or open orbit within the applied PN framework. A detailed examination of the software's performance is given in Ref. \cite{csizmadia}. The source release and binary packages supported both on x86 and x86\_64 platforms are available at the group's website \cite{CBwaves-web}.

In the PN formalism the spacetime is assumed to be split into the near and wave zones. The field equations for the perturbed Minkowski metric are solved in both regions. A fourth-order Runge--Kutta (RK4) method with adaptive step-size control is carred out to numerically solve for the 3PN-accurate near-field radiative dynamics at each time $t > t_{0}$, where $t_{0}$ is the time of arrival of the signal at the detector, while the far-zone radiation field \cite{kidder} decomposed as 
\begin{equation} \label{radiation-field-eq}
\begin{array}{ll}
h_{ij} = & \displaystyle \frac{2G\mu}{c^4 D}(Q_{ij} + P^{0.5}Q_{ij} + PQ_{ij} + PQ_{ij}^{\text{SO}} \\[10pt]
& + P^{1.5}Q_{ij} + P^{1.5}Q_{ij}^{\text{SO}} + P^{2}Q_{ij} + P^{2}Q_{ij}^{\text{SS}})
\end{array}
\end{equation}
is determined in harmonic coordinates by a simultaneous evaluation of orbital elements $(\phi,\, r,\, n)$ where $D$ is the distance (typically a few Mpc) to the GW source of that consists of two point particles of masses $m_{1}$ and $m_{2}$ and $\mu = m_{1}m_{2}/M$ is its reduced mass. The term $Q_{ij}$ is the Newtonian mass quadrupole moment, $P^{0.5}Q_{ij},\, P^{1.5}Q_{ij},\, P^{2}Q_{ij}$ are higher-order relativistic corrections up to 2PN order beyond the Newtonian term while  $PQ_{ij}^{\text{SO}},\, P^{1.5}Q_{ij}^{\text{SO}} ,\, P^{2}Q_{ij}^{\text{SS}}$ are corrections arising from spin--orbit and spin--spin effects, respectively. Here, for brevity, we will not repeat lengthy PN coefficients. They are written out explicitly in the appendix of Ref. \cite{csizmadia}.

\begin{figure}
	\begin{minipage}{.5\linewidth}
		\centering
		\subfloat[Integration run-times for equal-mass systems ($\bar{q}=0$).]{\label{fig:Tgen1}\includegraphics[scale=.65]{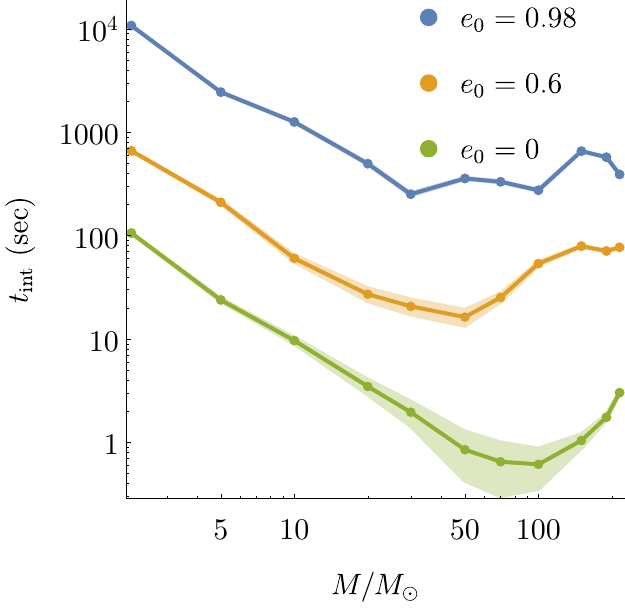}}
	\end{minipage}%
	\begin{minipage}{.5\linewidth}
		\centering
		\subfloat[Integration run-times for systems on circular orbit ($e_{0}=0$).]{\label{fig:Tgen2}\includegraphics[scale=.65]{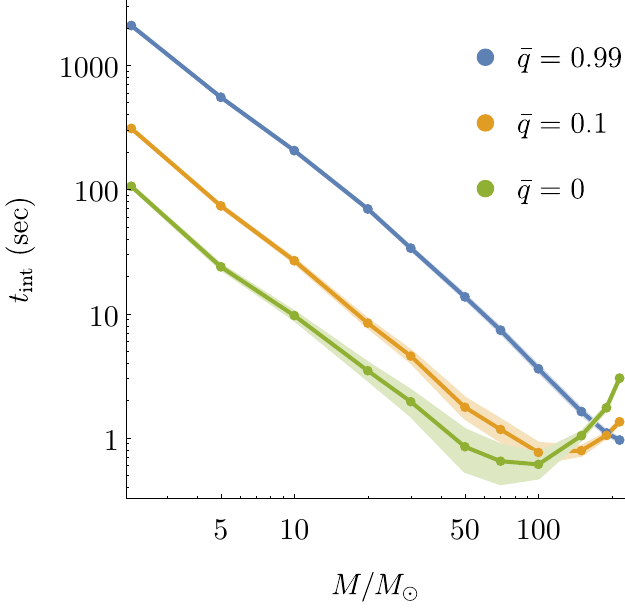}}
	\end{minipage}\par\medskip
	\caption{The integration run-time $t_{\text{int}}$ increases exponentially with decreasing total mass $M$. With increasing initial eccentricity $e_{0}$ (left panel) or mass disparity $\bar{q}$ (right panel), $t_{\text{int}}$ grows asymptotically at a significantly faster rate than with decreasing $M$. The integration time of those template waveforms that enter a detector's sensitivity band at a frequency of $10$ Hz has been measured $20$ times, each at $11$ distinct values of $M \in[2.15 M_{\odot},\, 215 M_{\odot}]$ for three distinct values of initial eccentricity; $e_{0}=\{0,\, 0.7,\, 0.98\}$ and mass disparity; $\bar{q}=\{0,\, 0.1,\, 0.99\}$ represented by blue, orange and green dots, respectively. The template waveforms were generated at a uniform sampling frequency $16.384$ kHz. Around each median curve of corresponding $t_{\text{int}}$ values, the shaded bands represent their respective $95\%$ point-wise confidence band.}
	\label{fig:Tgen}
\end{figure}

Radiative orbital dynamics involving all possible correction terms up the 3PN order beyond the Newtonian term are written out explicitly in terms of mean motion $n$ and orbital eccentricity $e$ in Ref. \cite{konigsdorffer}. The secular evolution is treated adiabatically, assuming that the timescales of the shrinkage of orbits (due to gravitational energy radiation $\dot{E}$) and the precession (due to angular momentum flux $\dot{J}$) are much longer than that of the orbital period. Consequently, the functions $(\dot{x}_{\text{1PN}},\, \dot{e}_{\text{1PN}}\ldots)$ in the equations derived from $\dot{E}$ and $\dot{J}$ depend only on the eccentricity $e$, and not on eccentric anomaly $u$. Hence, the adiabatic evolution equations for  $x \equiv (M\omega)^{2/3}$ and $e$ form a closed system, and can be solved independently of the Kepler's equation. Given initial conditions $x(0)$ and $e(0)$, we can solve the system of ordinary differential equations numerically to obtain $x(t)$ and $e(t)$. The integration of the equations of motion is terminated at the innermost stable circular orbit (ISCO), which is located at
\begin{equation} \label{ISCO-radius}
r_{\text{ISCO}} = 6GM/c^{2}
\end{equation}
in Schwarzschild spacetime (for a non-spinning NS/BH). The orbital angular frequency at the ISCO is
\begin{equation} \label{ISCO-frequency}
f_{\text{ISCO}} = c^3/(6\sqrt{6}\pi G M),
\end{equation}
which marks the end of the inspiral phase. Fig. \ref{fig:Tgen} demonstrates that the integration run-time $t_{\text{int}}$ depends sensitively both on the initial eccentricity and on the disparity of components' masses $(m_{1},\, m_{2})$ in a binary system. The $t_{\text{int}}$ increases exponentially with decreasing total mass $M$. The mass disparity, defined by $\bar{q} \equiv 1 - q$, allows better comparability with $e_{0}$ than $q$ itself, considering that $t_{\text{int}}$ asymptotically increases -- faster than with decreasing $M$ -- towards infinity as either $e_{0}$ (left panel) or $\bar{q}$ (right panel) tends to 1. The physical interpretation of these competing trends is very simple:
	\begin{enumerate}
		\item The lighter the components of the binary are, the longer it takes for them to gradually descend onto their ISCO through a sequence of increasingly circular orbits. \cite{pfeiffer}
		\item The more eccentric the orbit was initially, the longer it takes to shed its residual eccentricity over many orbital periods. \cite{peters}
		\item Among different configurations of equal total mass, the one with the largest mass disparity has the longest inspiral time for harbouring the lightest component. \cite{pfeiffer}
	\end{enumerate}
At the high total-mass region on Fig. \ref{fig:Tgen}, the influence of first trend grows comparable to that of the last two to reverse the trend of decreasing integration run-time. Fig. \ref{fig:parameter-space} shows the influence of $M$ and $q$ on the length of integration run-time $t_{\text{int}}$ from a different aspect. Excluding the red and yellow dots, each point in the coloured triangular region is assigned to a hue level running from dark to light as the value of $t_{\text{int}}$ increases on a logarithmic scale. The dark blue `basin' represents the region where $M$ and $q$ simultaneously lower the value of $t_{\text{int}}$ to its minimum. Isoclines running in parallel are connecting points at which $t_{\text{int}}$ has the same value, therefore they are associated with horizontal lines in Fig. \ref{fig:Tgen} (b). The influence of growing $q$ becoming comparable to and gradually greater than that of $M$ accounts also for the drift from the linear rising trend in the curvature of isoclines that occurs at the high-$q$ region on Fig. \ref{fig:parameter-space}. Although Fig. \ref{fig:parameter-space} suggests that over $85\%$ of the waveform templates of initial eccentricity $e_{0} = 0$ are computed up to $10$ seconds, in fact, only $4.6\%$ of all waveform templates require less then $10$ seconds to integrate, as demonstrated in Fig. \ref{fig:freqdist}. Out of a total of 1800, only those 120 templates are shown in Fig. \ref{fig:parameter-space} that are located in the $e_{0} = 0$ plane. Still, the figure illustrates well that in the same $e_{0}$-plane the frequency of templates with little $t_{\text{int}}$ is extremely high compared to that of templates with large $t_{\text{int}}$, regardless of the value $e_{0}$.\\
	
In the next section we shall give a quantitative description of the summary statistics computed from the relative frequency of occurrence (or empirical probability) of the integration time-runs.

\subsection{Probability distribution of integration run-times}\label{sec:probability-distribution}
Let $\mathbb{T} = \{t^{\text{int}}_{1},\, t^{\text{int}}_{2}, \ldots,\, t^{\text{int}}_{L}\}$ be a univariate independent and identically distributed (IID) finite data sample drawn from the probability (or relative frequency) distribution of the \emph{discrete} random variable $t \in \mathbb{T}$ while a discrete set of $L$ time-domain input waveforms
\begin{equation} \label{generated-waveform-set}
	\bm{h}(t) \equiv \{h(t; \lambda_{l})\}_{l=1}^{L}
\end{equation}
is computed at each parameter point $\lambda_{l}$ (see Sec. \ref{sec:parameter-space}) by evaluating eq. \eqref{radiation-field-eq} at a distance $D = \mu$ simultaneously with the integration of the equations of motions at 3PN order that requires integration run-times $t^{\text{int}}_{l}$. \\

\begin{figure}[h]
	\includegraphics[scale=0.80]{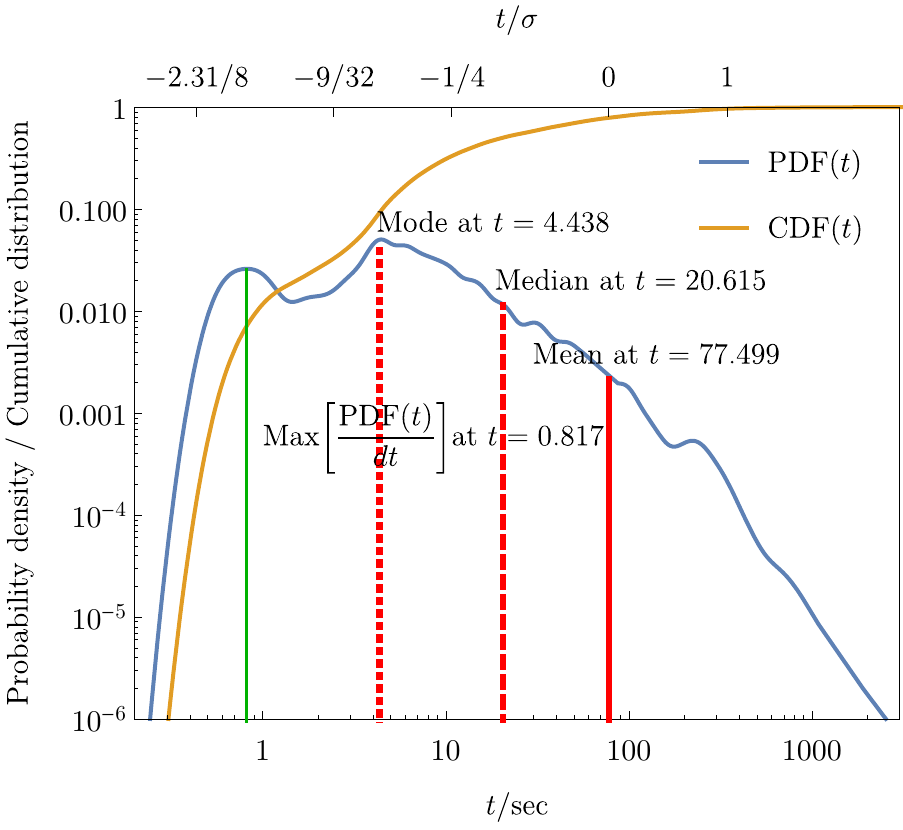}
	\caption{\label{fig:freqdist} 
		PDF denoted by $\mathscr{f}_{t}(t)$ (blue line) and CDF by $\mathscr{F}_{t}(t)$ (orange line) are displayed as functions of the random variable $t \in [\inf\mathbb{T},\, \sup\mathbb{T}]$, corresponding to $t_{\text{int}}$-values, which is measured in seconds on the lower horizontal axis and in standard deviation ($\sigma = 266.885$ sec) round the mean value of $t$ on the upper horizontal axis. The smooth KDE with adaptive bandwidth is based on the data sample $\mathbb{T}$ collected from the integration run-times of $L = 1800$ waveform templates that were generated in the parameter space $\Omega$, described in Sec. \ref{sec:parameter-space}. The location of the \emph{mode}, the \emph{median} and the \emph{arithmetic mean} are illustrated by dotted, dashed, and solid red lines, respectively in ascending order of their locations. This order of the measures of central tendency is a characteristic feature of right-skewed (positive skewness) distributions.}
\end{figure}

Since we do not make any prior assumption about the probability distribution, we shall use a non-parametric model where the \emph{statistical measures} are determined by the finite data sample $\mathbb{T}$. In statistics, kernel density estimation (KDE) is a fundamental data-smoothing technique that provides a non-parametric estimate, based on observed data $\mathbb{T}$, of an unobservable underlying probability density function (PDF) of the \emph{continuous} random variable $\inf\mathbb{T} \leq t \leq \sup\mathbb{T}$. A PDF, denoted by $\mathscr{f}_{t}$ and illustrated in Fig. \ref{fig:freqdist}, is a non-negative Lebesgue-integrable function that defines the cumulative distribution function (CDF) of a real-valued random variable $t$, evaluated at a value $t'$ as
\begin{equation} \label{probability-distribution}
\mathscr{F}_{t}[t'] \equiv \Pr[t \leq t'] = \int_{-\infty}^{t'}\mathscr{f}_{t}[\tau]d\tau.
\end{equation}
It represents the probability that the random variable $t$, with the expected value given by
\begin{equation} \label{expected-value}
E[t] = \int_{-\infty}^{\infty}t' d\mathscr{F}_{t}[t'],
\end{equation}
takes on a value less than or equal to $t'$ and its kernel density estimator is
\begin{equation}
\hat{\mathscr{f}}_{h}[t] = \frac{1}{Lh}\sum_{l=1}^{L}K\left[\frac{t-t^{\text{int}}_{l}}{h}\right]
\end{equation}
where $K \geq 0$ is a symmetric kernel with total integral normalized to unity and $h > 0$ is the bandwidth (or smoothing parameter). One might intuitively choose $h$ as small as the data sample $\mathbb{T}$ allows; however, there is always a trade-off between the bias of the estimator and its variance. Another option is the use of adaptive bandwidth kernel estimators in which the bandwidth changes as a function of $t$.

A specific quantitative measure of the probability distribution is the $n$-th moment 
\begin{equation} \label{moment}
\mu_{n} \equiv E[(t-c)^{n}]
\end{equation}
of the continuous random variable $t$ about some central value $c$ (e.g. the mean, denoted by $\mu$) where $E$ is the expected value of $t$ defined by eq. \eqref{expected-value}. The graphical representation of the most common measures of central tendency (mean, median, mode) is depicted on Fig. \ref{fig:freqdist} with solid, dashed and dotted red lines, respectively. The PDF rapidly increases with the random variable $t$ up to a point at $t = 0.81653$ sec. From then onwards this monotone increase slows down and eventually comes to a halt at $t = 4.438$ sec, which marks the \emph{mode}, i.e. the most frequent value in the distribution. The \emph{median} which represents the value separating the higher half of the probability distribution from the lower half is located at $t = 20.615$ sec. The \emph{mean} which represents the first moment of the PDF ($\mu \equiv \mu_{1}$ in eq. \eqref{moment}) is situated at $t = 77.499$ sec. 

The central tendency of distributions is typically contrasted with its dispersion that measures the extent to which a distribution stretched or squeezed. Common measures of statistical dispersion are the variance and standard deviation: The \emph{variance} of $t$ is the second central moment, given by \eqref{moment} as $\operatorname{Var}[t] \equiv E[(t-\mu)^{2}]$ and the \emph{standard deviation} is its square root, denoted by $\sigma$. For the given distribution $\sigma = 266.885$ sec. Finally, the shape (or asymmetry) of probability distributions is quantitatively measured by the third and fourth central moments, called \emph{skewness} and \emph{kurtosis} and denoted by $\operatorname{Skew}[t] \equiv E[(t-\mu)^{3}/\sigma^{3}] = 18.56$ and $\operatorname{Kurt}[t] \equiv E[(t-\mu)^{4}/\sigma^{4}] = 490.04$, respectively.

\section{Template placement and the frequency grid} \label{sec:grids}
We may now discuss the placement of a grid of TD waveform templates in a compact parameter space, followed by the generation of a sequence of frequency-domain (FD) templates on a common finely sampled uniform frequency grid. The TD waveforms are Fourier transformed and split into their amplitude and phase parts. These functions are accuretely represented on a sparse frequency grid with only $\mathcal{O}(10^4)$ nodes, with a sampling frequency recorded well above the Nyquist frequency of the shortest time-series in the template bank.

\subsection{Template placement in an associated 3-dimensional parameter space} \label{sec:parameter-space}
The set of input waveforms \eqref{generated-waveform-set} is computed by \emph{CBwaves}, described in Sec. \ref{sec:eccentric-waveform}, at corresponding parameter points 
\begin{equation}
\bm{\lambda} \equiv \{\lambda_{l} \in \Omega\}_{l=1}^{L}
\end{equation}
in a compact $p$-dimensional parameter space $\Omega \subset \mathbb{R}^{p}$ where $p$ is the number of model parameters. We restrict ourselves to a feasible 3-dimensional parameter space consisting of totally ordered one-dimenstional sets of values of corresponding model parameters $(m_{1},\, m_{2},\, e)$ that define a sparse grid of points
\begin{equation}
\begin{array}{lll}
\bm{\lambda} & \equiv & \hspace{-2pt} \bm{m}_{1} \otimes \bm{m}_{2} \otimes \bm{e} = \{(m_{i}, m_{j}, e_{k}):  \phantom{\}} \\[10pt]
&  & \hspace{-2pt} \phantom{\{} i \in [0, i_{\text{max}}]; j \in [i, i_{\text{max}}]; k \in [0, k_{\text{max}}]\}
\end{array}
\end{equation}
covering the desired parameter range in the particular model involved. Owing to the invariance of input waveforms under exchange of the components' masses $(m_{1},\, m_{2})$, the values of the 2-dimensional index pair $(i,\, j)$ are constrained to a triangular sub-region in the positive quadrants where $i \leq j$. Considering that the waveform templates span a 3-dimensional parameter space, each template is successively placed into a single vector \eqref{generated-waveform-set} as indexed by
\begin{equation} \label{parameter-position}
l \equiv \left[\left(i_{\text{max}} - \frac{i - 1}{2}\right) i + j\right]k_{\text{max}} + k + 1
\end{equation}
in the range of values $1 \leq l \leq L$. This flat index corresponds to the position of templates in the parameter space. The total number of templates in the set is then expressed as
\begin{equation}
L = \left[\left(2i_{\text{max}} + 1\right)^2 - 2\right]k_{\text{max}}/8 + 1.
\end{equation}
It is desirable to work with a dense grid of short waveforms encompassing the late inspiral phase to make a better coverage of the selected region of the parameter space. For the sake of simplicity, we sample at equidistant parameter combinations within the region. Nevertheless, using a template placement algorithm that is based on a template-space metric over the parameter space makes a far more efficient coverage. \cite{kalaghatgi,cokelaer} Generally, the algorithms that use geometrical techniques concentrate more points near the boundaries of the region and at lower mass-ratios.
\begin{figure}[h]
	\includegraphics[scale=0.80]{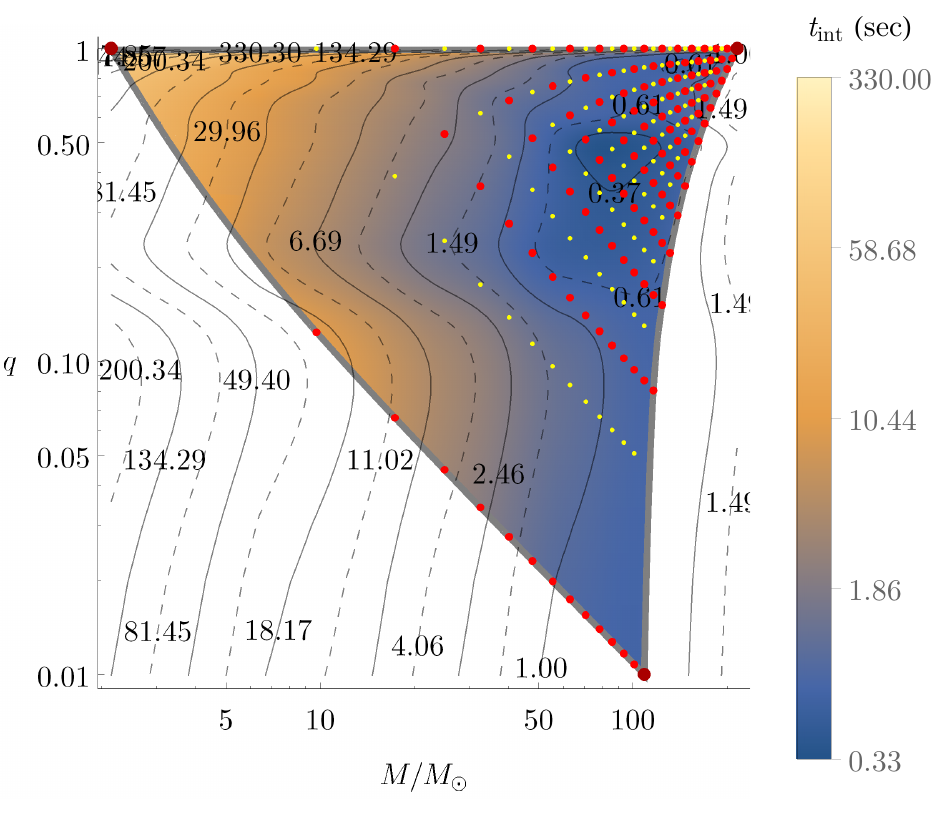}
	\caption{\label{fig:parameter-space} 
		The template bank $\bm{h}(t)$ of $L = 1800$ waveform templates was set up over a domain $\{(M,\, q,\, e_{0})|M \in [2.15M_{\odot}, 215M_{\odot}],\, q \in [0.01, 1],\, e_{0} \in [0, 0.95]\} \subset \Omega$ by computing eq. \eqref{generated-waveform-set} with uniform grid spacings $\{\Delta M = 14.19M_{\odot},\, \Delta q = 0.06,\, \Delta e_{0} = 0.063\}$. Red points, confined within a triangular region with a boundary $\partial\Omega$ (thick gray line), represet the parameter points of those $120$ input waveforms that are situated in the $k = 0$ plane section of the parameter space $\Omega$. In order to measure the accuracy of the ROM of waveforms, eq. \eqref{inerpolated-waveform} is evaluated at equidistant parameter points (yellow points) from their respective nearest basis-waveform neighbors. Each background point in the coloured triangular region is assigned to a hue level running from dark to light as the value of the integration run-time $t_{\text{int}}$ increases on a logarithmic scale. $t_{\text{int}}$ increases exponentially with decreasing total mass $M$ and grows asymptotically at a significantly faster rate with increasing mass disparity $\bar{q}$. The dark blue `basin', where the great majority of template waveforms are concentrated, represents the region where $M$ and $q$ simultaneously lower the value of $t_{\text{int}}$ to its minimum. Isoclines (thin gray curves) running in parallel are connecting points at which $t_{\text{int}}$ has the same value. A drift from the linear rise in the curvature of isoclines occurs at the high-$q$ region, where the influence of growing $q$ becomes comparable to and gradually greater than that of $M$.}
\end{figure}

The set of initial eccentricity $\{e_{k}: 1 \leq k \leq k_{\text{max}}\}$ is chosen to cover the entire interval $[0,\, 0.95]$ and the mass ratio $q \equiv m_{1}/m_{2} \leq 1$ is allowed to range between equal mass at $q = 1$ and relatively extreme systems at $q \approx 0.01$ with total mass $M/M_{\odot} \in [2.15, 215]$. In terms of the symmetric mass ratio $\eta = \mu/M$ the model covers approximately $\eta \in [0.01, 0.25]$. Fig. \ref{fig:parameter-space} shows the placement of those $L = 120$ templates (red dots) that are situated in the $k = 0$ plane section of the parameter space $\Omega$, out of a total of $1800$ templates and are collected in $\bm{h}(t)$. These templates are confined within a triangular region with a boundary $\partial\Omega$ (thick gray line).

\subsection{Production of frequency-domain waveforms} \label{sec:fourier-transform}
For optimal orientation all time-domain waveforms in eq. \eqref{generated-waveform-set}, composed of their two fundamental polarizations $h_{+}$ and $h_{\times}$ in the dominant $\mathscr{l} = \mathscr{m} = 2$ mode are represented by complex-valued GW strain amplitudes
\begin{equation} \label{complex-TDwaveforms}
h_{n}(\lambda_{l}) \equiv h_{+}(t_{n}; \lambda_{l}) - ih_{\times}(t_{n}; \lambda_{l})
\end{equation}
at $N$ equidistant grid points 
\begin{equation} \label{time-grid}
\{t_{n} = n\Delta t\}_{n \in [0, N-1]}
\end{equation}
as elements of a finite sequence of $N$ regularly spaced samples of the complex-valued TD waveforms $\{h_{0}(\lambda_{l}),\, h_{1}(\lambda_{l}),\, \ldots,\, h_{N-1}(\lambda_{l})\}$. The sequence is converted by a \emph{fast Fourier transform} (FFT), denoted by a linear operator $\mathscr{F}: h \to \tilde{h}$, into an other equivalent-length sequence of regularly spaced samples
\begin{equation} \label{FFT-definition}
\{\tilde{h}_{k}(\lambda_{l})\}_{k \in [-N/2, N/2-1]} = \mathscr{F}\{h_{n}(\lambda_{l})\}_{n \in [0, N-1]}
\end{equation}
evaluated at the same $N$ equidistant frequency grid points $\{f_{-N/2},\, \ldots,\, f_{0},\, \ldots,\, f_{N/2-1}\}$ considering that ROM construction, to be discussed in Sec. \ref{sec:rom}, will require a set of values that reside in the same grid points over all the waveforms in the template bank.
\begin{enumerate}
	\item This is achieved by having the length of all frequency series truncated to that of the shortest waveform in time, denoted by 
	\begin{equation} \label{time-length}
	T = t_{N-1} - t_{0}.
	\end{equation}	
	This particular waveform is associated with the highest mass, lowest eccentricity configuration $(i=i_{\text{max}},\, j =i_{\text{max}},\, k=0)$ in the template bank and its position in the parameter space, given by eq. \eqref{parameter-position}, is $l_{\text{short}} = (i_{\text{max}}+3)i_{\text{max}}k_{\text{max}}/2 + 1$. 
	\item Another possible way, used by \cite{purrer,purrer2}, to adjust the frequency series to the same length is to make the shorter-length waveforms of sufficient length by extending them with other templates such as TaylorF2.
\end{enumerate}
The Fourier coefficients in eq. \eqref{FFT-definition}, given by
\begin{equation} \label{Fourier-coefficients}
\displaystyle \tilde{h}_{k}(\lambda_{l}) \equiv \sum _{n=0}^{N-1}h_{n}(\lambda_{l}) e^{-2\pi i kn/N}, \quad k \in [0, N-1],
\end{equation}
are complex-valued functions of the frequency $f_{k}$ which encodes both the amplitude and the phase,
\begin{equation} \label{FFT-amplitude-and-phase}
\begin{array}{lll}
\tilde{h}^{(A)}_{kl} \hspace{-2pt} & = & \hspace{-2pt} \sqrt{\operatorname{Re}[\tilde{h}_{k}(\lambda_{l})]^2 + \operatorname{Im}[\tilde{h}_{k}(\lambda_{l})]^2}/N, \\[10pt]
\tilde{h}^{(\phi)}_{kl} \hspace{-2pt} & = & \hspace{-2pt} -i\ln\left(\tilde{h}_{k}(\lambda_{l})/|\tilde{h}_{k}(\lambda_{l})|\right),
\end{array}
\end{equation}
respectively. In this interpretation, $\tilde{h}_{k}(\lambda_{l})$ corresponds to the cross-correlation of the time sequence $h_{n}(\lambda_{l})$ and an $N$-periodic complex sinusoid $e^{2\pi i kn/N}$ at a frequency point $f_{k} \equiv k/N$ that represents $k$ cycles of the sinusoid. Therefore, eq. \eqref{Fourier-coefficients} acts in place of a \emph{matched filter} for that frequency. Now, the sequence of frequency-domain waveforms \eqref{FFT-definition} can be re-expressed as `chirps' in a simple form
\begin{equation} \label{FDwaveforms}
\{\tilde{h}_{k}(\lambda_{l})\} = \{\tilde{h}^{(A)}_{kl}\exp(i\Lambda\tilde{h}^{(\phi)}_{kl})\}
\end{equation}
where the oscillation degree $\Lambda$ is a large number. The behavior of GWs in the late inspiral phase is highly oscillatory, but the amplitude and the phase themselves are smoothly varying functions of frequency. \cite{candes} It will thus be more expedient to perform high-accuracy parametric fits of the phase and amplitude given by \eqref{FFT-amplitude-and-phase} rather than of the complex waveform \eqref{complex-TDwaveforms} itself. The preprocessed amplitudes and phases are collected in the columns of separate template matrices $\{\mathcal{H}^{(A)},\, \mathcal{H}^{(\phi)}\} \in \mathbb{R}^{N \times L}$,
\begin{equation} \label{template-matrix}
\mathcal{H} = (\tilde{h})_{kl} \in \mathbb{C}^{N \times L}
\end{equation}
where we have droped the amplitude or phase labels for brevity and where $L$ is the total number of templates, and each template $\tilde{h}_{l}(f_{k})$ is given on a common freqency grid of length $N$. We may choose to represent the waveforms at a large number of frequency points so that  $N \gtrsim L$.
\begin{figure}[h]
	\includegraphics[scale=0.80]{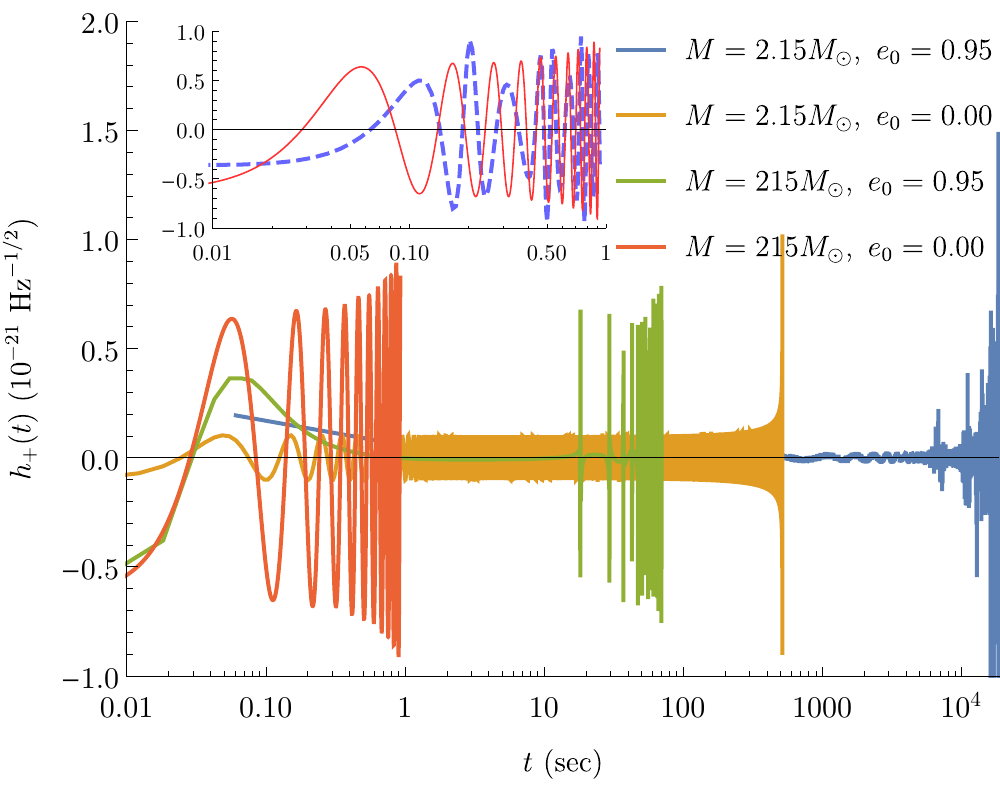}
	\caption{\label{fig:TDwaveforms} 
		The panels illustrate the inspiral evolution of equal-mass BBH/BNS systems ($\bar{q}=0$) starting at a Keplerian mean orbital frequency of $5$ Hz at a distance $D = \mu$. The four most distinct template waveforms were generated by \emph{CBwaves} at a uniform sampling frequency of $16.384$ kHz with extreme values of total mass $M = \{2.15 M_{\odot},\, 215 M_{\odot}\}$ and initial eccentricity $e_{0}=\{0.00,\, 0.95\}$ in the investigated parameter space $\Omega$. (See large red points on Fig. \ref{fig:parameter-space}.) The top inset panel presents the last $N = 15,000$ points of the longest waveform (blue) projected onto an equal number of points of the shortest waveform (red).}
\end{figure}

\subsection{Definition of a regularly spaced high-resolution frequency grid} \label{sec:frequency-grid}
Provided that $T$ in \eqref{time-length} is the longest time length, the time spacing is defined as
\begin{equation} \label{time-spacing}
\Delta t = T/(N-1)
\end{equation}
by eqs. \eqref{time-grid} and \eqref{time-length}. The time spacing and the number of time steps $N$ in the grid \eqref{time-grid} are chosen such that the FD waveforms \eqref{FDwaveforms} are sampled at a rate of $f_{\text{s}}$ and cover a suitable and well-resolved frequency range $[f_{\text{low}},f_{\text{high}}]$. 
\begin{enumerate}
	\item The lower limit of the frequency range $f_{\text{low}}$ is specified by the low-frequency cutoff of the detector noise spectrum which is close $f_{\text{cutoff}} = 10$ Hz for advanced detectors design.
	\item The upper limit $f_{\text{high}}$ is determined to be at $f_{\text{ISCO}} = 2.045$ kHz by the ISCO frequency \eqref{ISCO-frequency} for the lowest total mass configuration of interest $M = 2.15 M_{\odot}$.
\end{enumerate}

The Nyquist criterion requires the sampling frequency to be at least twice the highest frequency contained in the signal to avoid aliasing. Thus, the smallest sufficient sampling frequency is $f_{\text{s}} = 4096$ samples per second for being the first power of 2 to meet the criterion. Note that the typical sampling rate being used by aLIGO and aVirgo observatories in ongoing searches for GWs is at 2048 Hz. \cite{LIGO} Instead, an equidistant grid with $N = 4000$ grid points is sampled at $f_{\text{s}} = 16.384$ kHz in the frequency band $Mf \in [0.0001,0.0216]$ in geometric units $(G = c = 1)$. The total mass $M$ is expressed in units of geometrized solar mass by $M_{\odot}[\text{s}] = G/c^3 \times M_{\odot}[\text{kg}] \approx 4.93 \times 10^{-6}$ sec. At the time resolution $\Delta t = 1/f_{\text{s}} \approx 4.59 M$ which corresponds to a Nyquist frequency 
\begin{equation} \label{Nyquist-frequency}
f_{\text{Ny}} = f_{\text{s}}/2 \approx 2.03 \times 10^{-32} M^{-1},
\end{equation}
a waveform long enough for the BNS system of total mass $M = 2.15 M_{\odot}$ down to $f_{\text{low}} = 2.48 \times 10^{-35} M^{-1}$ is given and is about $T = (N-1)\Delta t \approx 1.83 \times 10^{4} M$ long in time. The spacing in frequency domain is 
\begin{equation} \label{frequency-spacing}
\Delta f = 2f_{\text{Ny}}/N,
\end{equation}
so the power will be either in positive or negative frequencies, depending on conventions and we need to consider only half of the FFT. Combining this with the relations (\ref{time-spacing}--\ref{Nyquist-frequency}), one has
\begin{equation} \label{Frequency-spacing}
\Delta f = \frac{N-1}{N}\frac{1}{T} \approx 5.45 \times 10^{-5} M^{-1}.
\end{equation}
Only half of the points in the FFT spectrum are unique, the rest are symmetrically redundant. Thus, the points of negative frequencies contain no new information on the periodicity of the random number sequence. Which amounts to swapping the left and right half of the result of the transform.

\section{SVD-based reduced-order surrogate model building} \label{sec:rom}
In this section we summarize some of the characteristic features of SVD that are especially useful for reduced-order modeling and discuss our approach to construct a compressed approximate representation of a collection of fiducial waveforms at the cost of truncation error. Next, projection coefficients of the waveforms are determined in terms of the reduced basis. In conclusion, the ROM is assembled from the reduced basis and projection coefficients interpolated over the parameter space. Our procedure follows the well-established strategy that has been pursued by P\"{u}rrer and by Cannon for building frequency-domain ROMs. \cite{purrer,purrer2,cannon,cannon2}

\subsection{Singular values and truncation error} \label{sec:SVD}
\begin{figure}[h]
	\includegraphics[scale=0.65]{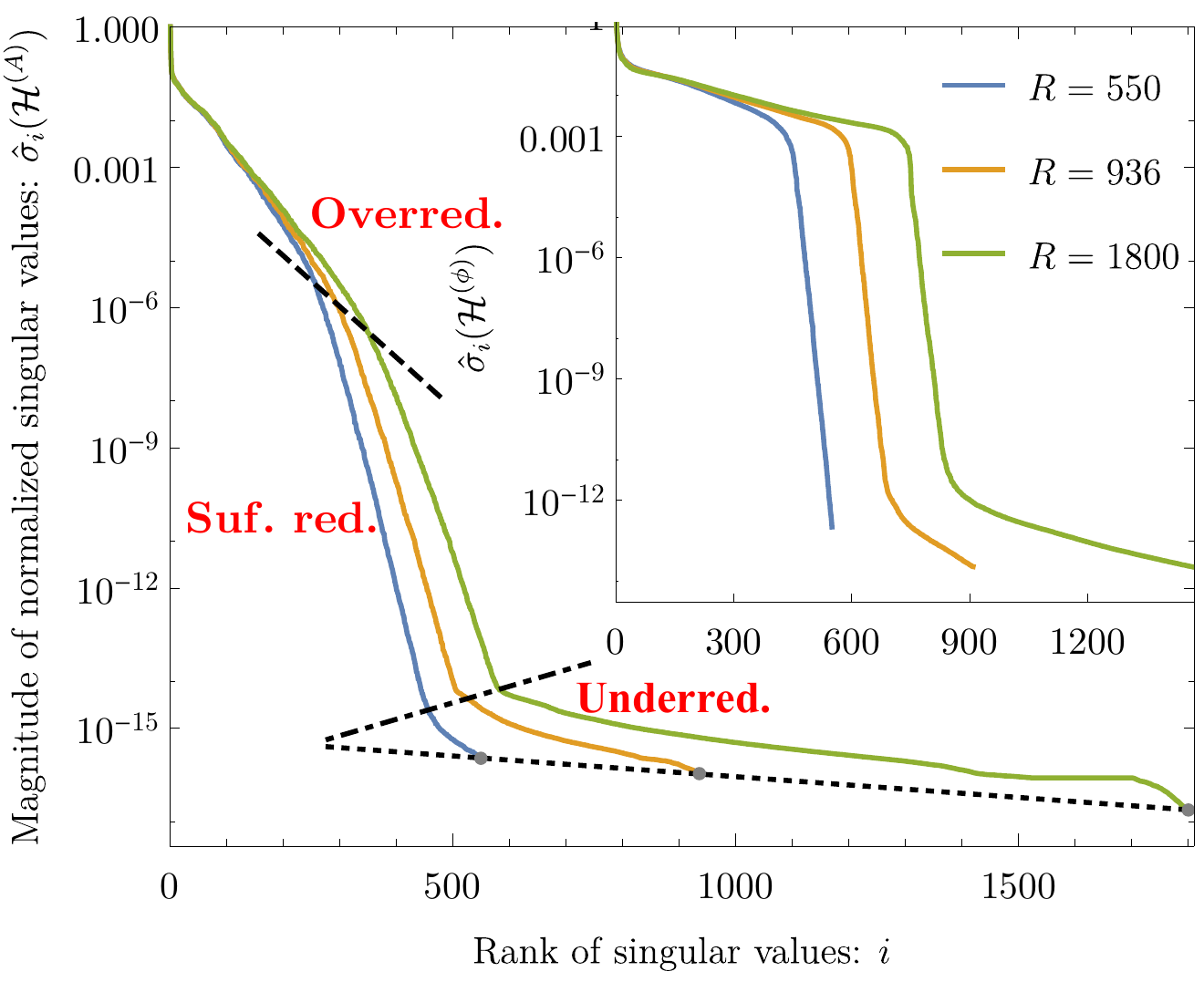}
	\caption{\label{fig:sigma}
		Normalized singular-value spectra of the template matrix for full ranks $R = \{550,\, 936,\, 1800\}$ are illustrated by blue, orange and green colour, respectively. 
		The horizontal axis represents the index of SVs, while the vertical axis represents the relative variance of SVs. The main panel displays the relative variance of $\hat{\sigma}_{i}$ of the matrix $\mathcal{H}^{(A)}$ which encodes the amplitude part of waveform templates while the corresponding relative variance of $\hat{\sigma}_{i}$ for the phase is shown in the top inset. At $r = R - 1$ its infimum, $\hat{\sigma}_{r}$, falls onto a dotted black line given by $\log\hat{\sigma}_{r} - \log\sigma_{1}\approx -34.8877 - 0.00204394 R$. The rate at which the ratio decreases is significantly lower under the dashed black line given by $-6.23703 - 0.0250683 R$. Excluding waveforms in the lower section causes less errors by a magnitude much smaller than in the upper section.}
\end{figure}

Formally, the decomposition of the template matrix $\mathcal{H} \in \mathbb{C}^{N \times L}$ in eq. \eqref{template-matrix} is expressed by a factorization of the form
\begin{equation} \label{SVD-definition}
\mathcal{H} = V\Sigma U^{\dagger},
\end{equation}
where the complex unitary matrices 
\begin{equation}
\begin{array}{ll}
V & = [\bm{v}_{1}| \ldots |\bm{v}_{L}] \in \mathbb{R}^{L \times L}, \\
U & = [\bm{u}_{1}| \ldots |\bm{u}_{N}] \in \mathbb{R}^{N \times N}
\end{array}
\end{equation}
are orthogonal sets of non-zero eigenvectors of the non-negative self-adjoint operators $\mathcal{H}^{\dagger}\mathcal{H}$ and $\mathcal{H}\mathcal{H}^{\dagger}$ so that $U^{\dagger}U = \mathbb{I}$ and $V^{\dagger}V = \mathbb{I}$. The rank--nullity theorem states that the SVD \eqref{SVD-definition} provides a decomposition of the range of $\mathcal{H}$. \cite{purrer} Accordingly, the left-singular vectors (or \emph{eigensamples}) $\{\bm{v}_{i} \in V\}$ provide an orthonormal basis
\begin{equation} \label{Graham-Smith}
\operatorname{range}(\mathcal{H}) = \operatorname{span}\{\bm{v}_{1},\ldots, \bm{v}_{R}\}
\end{equation}
for the range of $\mathcal{H}$ (columnn space) where the maximal number of linearly independent columns of $\mathcal{H}$ is $R \equiv \operatorname{rank}(\mathcal{H}) \leq L$. In a qualitative sense, each $\bm{v}_{i}$ represents a typical waveform pattern. The right-singular vectors (or \emph{eigenfeatures}) $\{\bm{u}_{i} \in U\}$ provide a basis for the domain of $\mathcal{H}$ (row space) and represent the evolution of the magnitude of each waveform along the frequency gridpoints. The diagonal entries of the rectangular matrix $\Sigma \in \mathbb{R}^{N \times L}$ correspond to the non-negative real \emph{singular values} (SVs) $\sigma_{1} \geq \ldots \geq \sigma_{s} \geq 0$ where $s = \min(N, L)$. SVs are roots of eigenvalues of $\mathcal{H}^{\dagger}\mathcal{H}$ (and of $\mathcal{H}\mathcal{H}^{\dagger}$) describing the spectrum of the template matrix $\mathcal{H}$, arranged in monotonically decreasing order (see Fig. \ref{fig:sigma}). If the number of frequency points is significantly larger than the number of waveforms (i.e. $L \ll N$) then a \emph{thin} SVD is a more compact and `economical' factorization of eq. \eqref{SVD-definition} than the \emph{full-rank} SVD that comprizes all $R$ eigensamples. In practice, low-rank matrices are often contaminated by errors, and for that reason they feature an effective rank $R_{\text{eff}}$ smaller than its exact rank $R$. The \emph{reduced-rank} approximation of the template matrix $\mathcal{H}$ is expressed by
\begin{equation} \label{reduced-SVD}
\mathcal{H}_{r} = \sum_{i=1}^{r}\sigma_{i}\bm{v}_{i}\otimes \bm{u}^{T}_{i}
\end{equation}
which comprises only those $r < R$ singular vectors which correspond to singular values of a significant magnitude. The approximated representation \eqref{reduced-SVD} of the fiducial template bank $\mathcal{H}$ is the $r$-th partial sum of the outer-product expansion of the expression \eqref{SVD-definition} where $r$ denotes the desired \emph{target rank}. The Eckart--Young theorem \cite{eckart} implies that the low-rank SVD in eq. \eqref{reduced-SVD} provides the optimal rank-$r$ reconstruction of the template matrix
\begin{equation}
\mathcal{H}_{r} \equiv \underset{\operatorname{rank}(\mathcal{H}'_{r}) = r}{\operatorname{argmin}}\parallel\mathcal{H} - \mathcal{H}'_{r}\parallel
\end{equation}
in the least-square sense where the truncation error of approximated representation \eqref{reduced-SVD} in both the spectral and Frobenius norm is given by
\begin{equation} \label{Frobenius_truncation_error}
\begin{array}{ll}
\parallel\mathcal{H} - \mathcal{H}_{r}\parallel_{2} & = \sigma_{r+1}(\mathcal{H}), \\
\parallel\mathcal{H} - \mathcal{H}_{r}\parallel_{F} & = \sqrt{\sum^{\min(N, L)}_{i = r+1}\sigma_{i}^{2}(\mathcal{H})},
\end{array}
\end{equation}
respectively. 

Fig. \ref{fig:sigma} shows $\hat{\sigma}_{i} \equiv \sigma_{i}/\sigma_{1}$ on logarithmic scale as a function of the number of SVD components $i = \{1, \ldots, R\}$ involved in the approximated representation. Each $\hat{\sigma}_{i}$, which describes the relative magnitudes of the corresponding eigenfeatures, is computed from the truncated SVD \eqref{reduced-SVD} of template matrices with three distinct full-ranks $R = \{550,\, 936,\, 1800\}$ (i.e. total number of templates). The truncation error in the approximation, in accordance with eq. \eqref{Frobenius_truncation_error}, decreases with the number of SVD components retained. The ultimate accuracy (or minimal error) achievable is limited by the total number of templates $L$ that the original template matrix $\mathcal{H}$ contains. The growing rate of decay in the SV spectrum demonstrates that the individual SVD components gradually lose their relevance for being included in the approximation. In this respect, the spectrum has three clearly distinctive regions characterized by the rate at which SVs decrease:
\begin{enumerate}
	\item \emph{Overreduced} SVDs $(k \lesssim 400)$ retain insufficient amount of information to construct a representation by the orthonormalization \eqref{Graham-Smith} with less than relative error of $10^{-5} - 10^{-6}$. The initial steep exponential fall attests that the information contained in the corresponding eigenfeatures is predominantly relevant. In fact, the first few components shown on Fig. \ref{fig:BasisFunc} contain roughly $90\%$ of all the information on the input waveforms, regardless of $\operatorname{rank}(\mathcal{H})$. Then, SVs decrease at a much lower, yet a slowly increasing rate, practically indistinguishable for different values of full rank $R$.
	\item \emph{Sufficiently reduced} SVDs  $(400 \lesssim k \lesssim 500-600)$ efficiently select the relevant information, so that the relative error of representation \eqref{relative-error} is kept well-suppressed while the number of SVD components stored in the reduced-rank template matrix is significantly lower than that of the full-rank. The larger the full-rank $R$ is, the more SVD components have to be kept to achieve the same accuracy of representation.
	\item \emph{Underreduced} SVDs $(k \gtrsim 500-600)$ admit the lowest possible truncation errors, limited only by the numerical errors of the full-rank approximation itelf. However, the accuracy of reconstructed waveform representation improves at a rate much lower than in the preceding regions. The loss of relevant information content due to the reduction of the number of SVD components is inefficiently low compared to the improvement of accuracy. 
\end{enumerate}
Choosing an optimal target rank $r$ is highly dependent on the objective. One either desires a highly accurate reconstruction of the fiducial waveform templates, or a very low dimensional representation of the fundamental features in the templates. In the former case $r$ should be chosen close to the \emph{effective rank}, while in the latter case $r$ might be chosen to be much smaller. Fig. \ref{fig:sigma} demonstrates that choosing a target rank $r = 456$ for the smallest among fiducial template matrices will result in a truncation error related to $\hat{\sigma} = 2.66 \times 10^{-15}$ at $r = 456$.

\subsection{Assembly of the surrogate model} \label{sec:surrogate}
\begin{figure}
	\begin{minipage}{.4\linewidth}
		\centering
		\subfloat[Basis functions for the amplitude modes.]{\label{fig:BasisFuncA}\includegraphics[scale=.9]{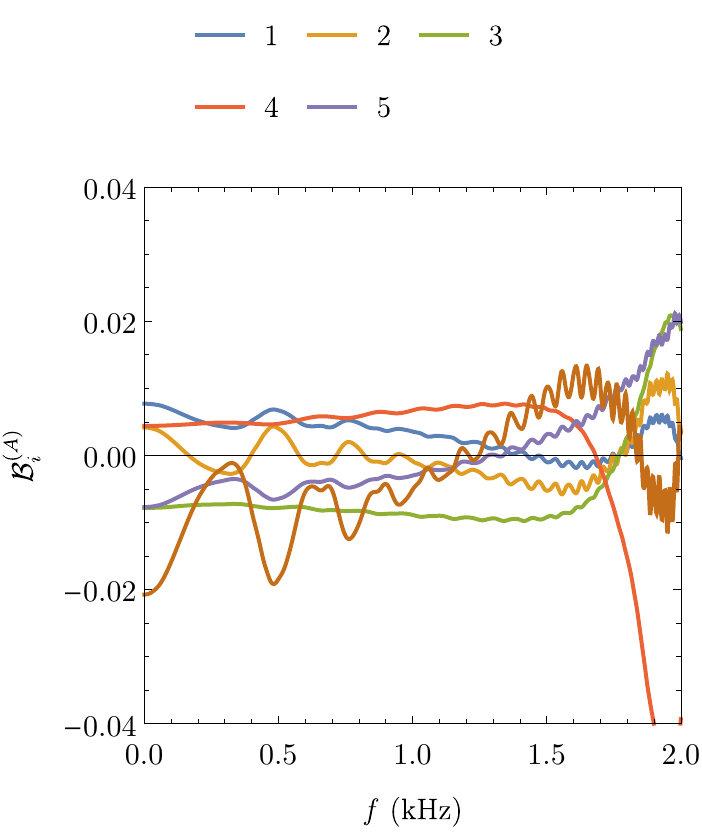}}
	\end{minipage}%
	\begin{minipage}{.4\linewidth}
		\centering
		\subfloat[Basis functions for the phase modes.]{\label{fig:BasisFuncPhi}\includegraphics[scale=.9]{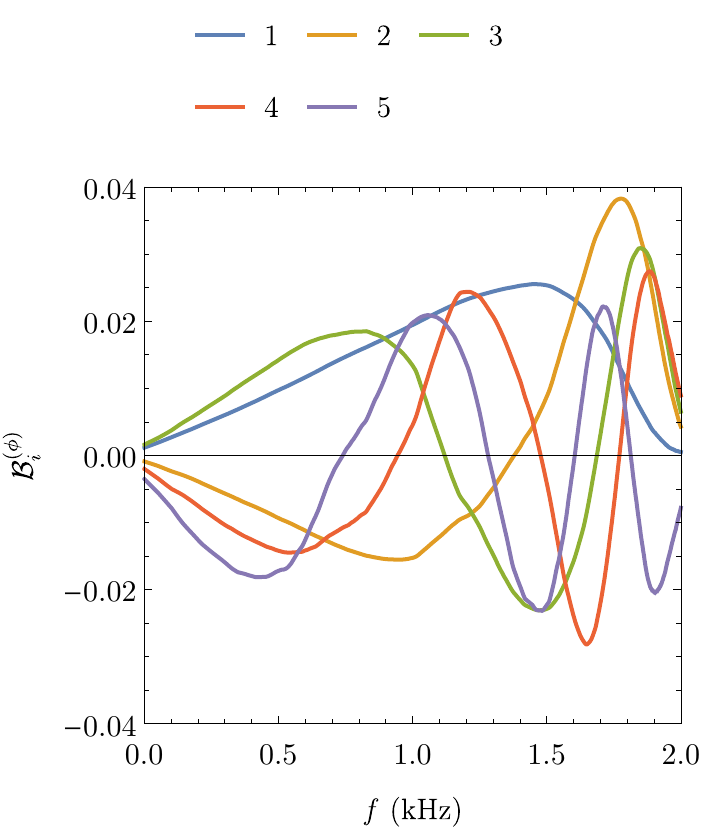}}
	\end{minipage}\par\medskip
	\caption{Reduced basis functions for the first 5 amplitude and phase SVD modes are represented at $N = 4000$ grid points in the frequency domain. The basis functions become increasingly oscillatory as their index $i$ increases.}
	\label{fig:BasisFunc}
\end{figure}

\begin{figure}[h]
	\includegraphics[scale=0.65]{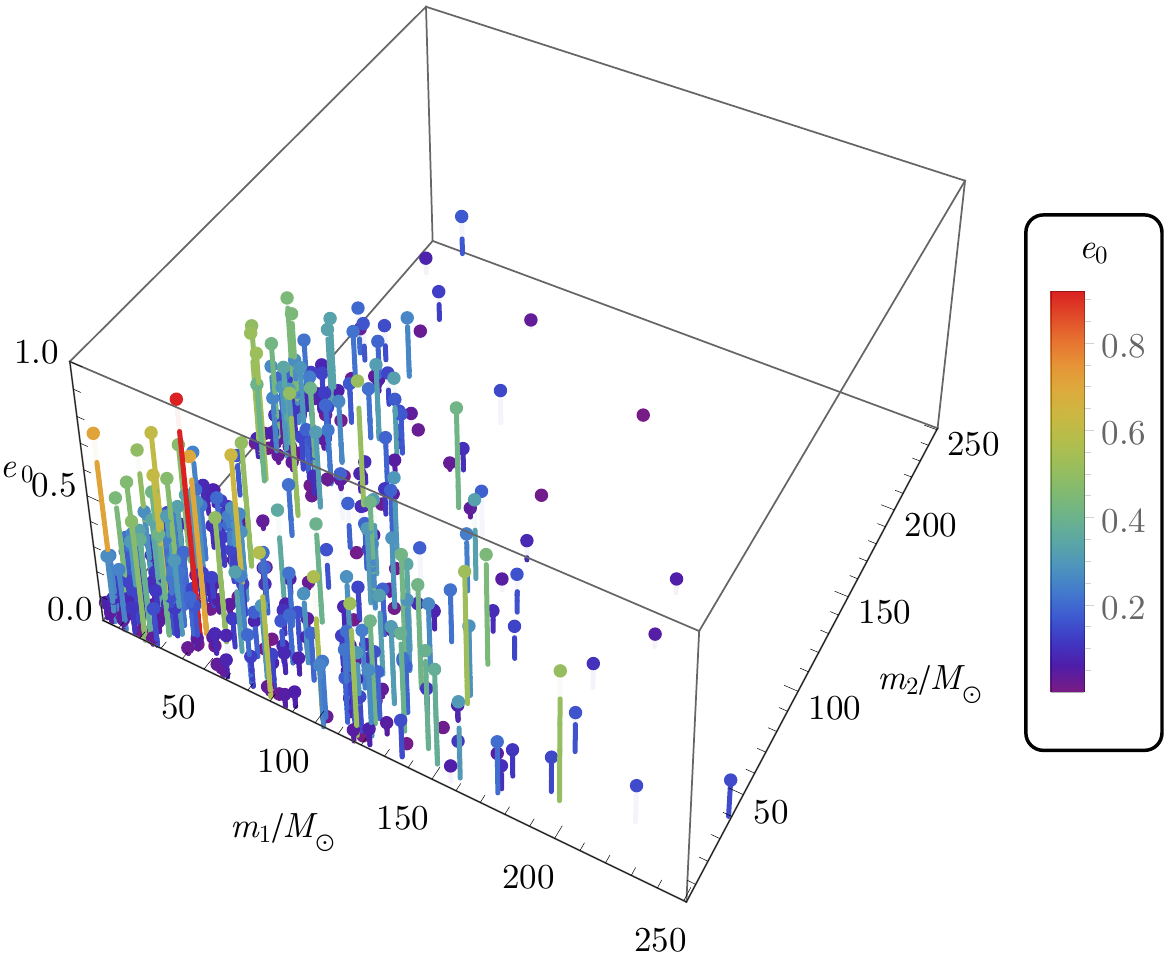}
	\caption{\label{fig:parameterspace}
		The SVD-based reduced-basis parameter choices in the 3-dimensional parameter space $(m_{1},\, m_{2},\, e_{0})$. Comparing the positions of the retained $r = 600$ templates to the placement of the original $R = 1800$ template shown in Fig. \ref{fig:parameter-space}, it becomes clear that primarily those parameters are selected that are associated with low-mass and low-eccentricity systems. Only a small fraction of the whole volume of the parameter space is covered.}
\end{figure}

The basis for the amplitude or phase space is given in the columns $\mathcal{B}_{i}$ of the matrix
\begin{equation}
\mathcal{B} \equiv
\begin{cases}
V_{L} \in \mathbb{R}^{N \times L},& \text{if } N > L \\
V \in \mathbb{R}^{N \times N},& \text{if } N \leq L
\end{cases}
\end{equation}
and a full-rank basis is desired. If $N < L$, then the information from $L$ waveforms at $N$ grid points is contained in a basis of dimension $N$. The reduced basis waveforms only resemble the physical behavior of frequency domain amplitudes and phases for the first basis function, the higher basis functions are oscillatory (see Fig. \ref{fig:BasisFunc}). To compress the model, a \emph{reduced basis} of rank $r$ is selected from the full-rank basis \eqref{Graham-Smith} in the form
\begin{equation}
\mathcal{B}_{r} = V_{r} = [\bm{v}_{1}| \ldots |\bm{v}_{r}] \in \mathbb{R}^{N \times r} \text{ for } r < R \leq N.
\end{equation}
For any $r$ the columns of $V_{r}$ are an optimal orthonormal bases for the starting waveforms. Notice that $\mathcal{B}_{r} \subset \mathcal{B}_{r+1}$, which demonstrates the underlying hierarchical nature of the generated template banks. \cite{purrer} Fig. \ref{fig:parameterspace} may serve as an illustration of the underlying sparsity of the selected basis in the parameter space. The identification of parameter values associated with the basis waveforms selected by SVD from the full template bank is not that straightforward as a greedy algorithm would pick values that parameters take. Nevertheless, it may safely be said that a very small part of the parameter space volume is covered; the parameter points are heavily concentrated at low-mass and low-eccentricity values. 

Hereafter the label $r$ on the rank-$r$ reduced basis will be dropped for brevity. Given the reduced bases $\mathcal{B}^{(A)}$ and $\mathcal{B}^{(\phi)}$, we compute projection coefficient vectors $\vec{\mu}$ for any given input waveform $\tilde{h} \in \mathbb{R}^{N}$ as follows
\begin{equation} \label{projection-coefficient}
\vec{\mu}(\tilde{h}) \equiv \mathcal{B}^{T}\tilde{h} \in \mathbb{R}^{r},
\end{equation}
where the labels referring to amplitude or phase were dropped for brevity. The projection coefficient vectors for all waveform templates are packed in the matrices $\mathcal{M}^{(A)}$ and $\mathcal{M}^{(\phi)}$ with entries
\begin{equation}
\mathcal{M}_{kl} = \mu_{k}(\tilde{h}_{l}) = (\mathcal{B}^{T}\mathcal{H})_{kl} \in \mathbb{R}^{r \times L}.
\end{equation}
Comparing with eq. \eqref{SVD-definition} we see that $\mathcal{M} = \mathcal{B}^{T}\mathcal{H} = -\Sigma U^{T}$ for a full-rank basis $\mathcal{B} = V$. It follows that the projection coefficient matrices are ordered in the same way as the individual waveforms in $\mathcal{H}$. To undo the packing of the waveforms in the matrices $\mathcal{M}$ we just partition the linear index $l$ that enumerates the waveforms in $\mathcal{H}$ and obtain a tensor
\begin{equation}
\mathcal{M}_{k,l_{q},l_{e}} = \mu_{k}(\tilde{h}_{(l_{q},l_{e})}) \in \mathbb{R}^{r \times L_{q} \times L_{e}}.
\end{equation}
To complete the model we define the projection coefficient vectors at any position in the chosen parameter space by suitable interpolants $I[\mathcal{M}](\lambda) \in \mathbb{R}^{r}$ for the amplitude and phase coefficient tensors $\mathcal{M}^{(A)},\, \mathcal{M}^{(\phi)}$. For each input waveform we have two corresponding $r$-vectors of projection coefficients (for amplitude and phase) that are interpolated over the parameter space. The frequency-domain ROM representation of waveform templates is then constructed in the form
\begin{equation} \label{inerpolated-waveform}
\begin{array}{lll}
\tilde{h}_{\text{S}}(\lambda; f) \hspace{-2pt} & \equiv & \hspace{-2pt} A_{0}(\lambda)I_{f}[\mathcal{B}^{(A)} \cdot I[\mathcal{M}^{(A)}](\lambda)] \\[8pt]
& & \hspace{-2pt} \times \exp\{i I_{f}[\mathcal{B}^{(\Phi)} \cdot I[\mathcal{M}^{(\Phi)}](\lambda)]\},
\end{array}
\end{equation}
where $\cdot$ denotes matrix multiplication, $I_{f}[\cdot]$ interpolates vectors in frequency on a suitable grid, and $A_{0}(\lambda)$ is an amplitude prefactor which is stored before the SVD takes place and an interpolant is computed over the parameter space.

\section{Accuracy and speedup for surrogate model predictions} \label{sec:efficiency}
Once a ROM is built, any surrogate waveform can be evaluated as a sum of reduced basis elements with incremental errors within the parameter range covered in the particular model. The main criteria for a successful ROM are that it facilitates data analysis applications that were infeasible with the fiducial waveform model and that it represents waveforms accurately. \cite{purrer2} This section is dedicated to appraise the overall performance of the ROM building discussed in Sec. \ref{sec:rom}. The first part of this section assesses the accuracy of surrogate model predictions in terms of the match between the surrogate model and the fiducial model. In the second part we provide an overview of the computational efficiency of the ROM with respect to computational complexity and cost relative to the cost of the fiducial model.
\begin{figure}[h]
	\includegraphics[scale=0.90]{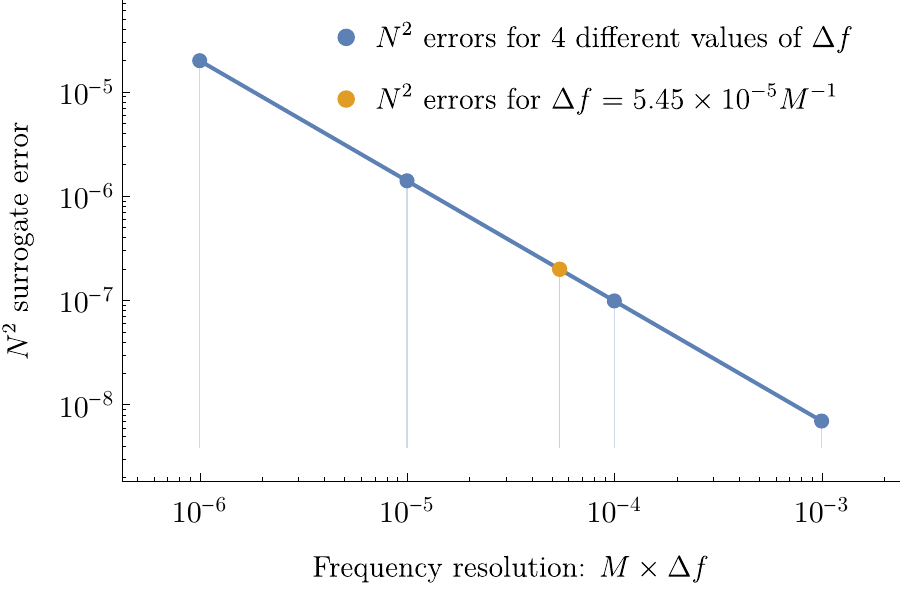}
	\caption{\label{fig:surr-error} 
		The linear trend in the change of surrogate error \eqref{surr-error} as a function of the resolution of the frequency grid. Higher resolution of sampling times (i.e. lower resolution for sampling frequencies) result lesser uncertainty in estimating the amplitudes and phases. Surrogate error $\Delta\tilde{h}^{2} = 1.98 \times 10^{-7}$ is marked with an orange point for a frequency-grid spacing $\Delta f = 5.45 \times 10^{-5} M^{-1}$ which was obtained in eq. \eqref{Frequency-spacing}. The value of surrogate error corresponds to the mean relative error of the amplitude $\Delta\tilde{h}^{(A)} \approx 4 \times 10^{-14}$ shown in Fig. \ref{fig:pointwise-error}.}
\end{figure}

\subsection{Reconstruction errors}
The overlap integral of two normalized waveforms, say, of a fiducial \emph{CBwaves} waveform $h_{\text{CB}}$ and its surrogate model prediction $h_{\text{S}}$, is given by the \emph{mismatch} (or \emph{unfaithfulness}) between the two waveforms and is defined as the normalized inner product \eqref{inner-product} maximized over time and phase shifts
\begin{equation}
\mathscr{M} \equiv 1 - \max_{t_{0},\phi_{0}}\frac{\langle h_{\text{CB}}, h_{\text{S}} \rangle}{\lVert h_{\text{CB}} \rVert \lVert h_{\text{S}}\rVert}
\end{equation}
with an inherited norm given by $\lVert h \rVert^{2} \equiv \langle h, h \rangle$. A natural inner product between the two waveforms is given by the complex scalar product
\begin{equation} \label{inner-product}
\langle \tilde{h}_{\text{CB}}, \tilde{h}_{\text{S}} \rangle \equiv 4\operatorname{Re}\int^{f_{\text{high}}}_{f_{\text{low}}}\frac{\tilde{h}_{\text{CB}}(f) \tilde{h}_{\text{S}}^{*}(f)}{S_{\tilde{h}}(f)}df
\end{equation}
where the tilde denotes Fourier transformation given in eq. \eqref{FFT-definition}, $\tilde{h}_{\text{S}}^{*}(f)$ is the complex conjugate of $\tilde{h}_{\text{S}}(f)$, $S_{\tilde{h}}(f)$ is the one-sided power spectral density (PSD) of the detector noise and $f_{\text{low}}$, $f_{\text{high}}$ are suitable cutoff frequencies for detector sensitivity. The low-frequency cutoff depends on the PSD and is at $10$ Hz for advanced detectors design. The high-frequency cutoff is at $2.045$ kHz, which is the ISCO frequency of the lowest total-mass configuration in our fiducial model, discussed in Sec. \ref{sec:frequency-grid}.
\begin{figure}[h]
	\includegraphics[scale=0.90]{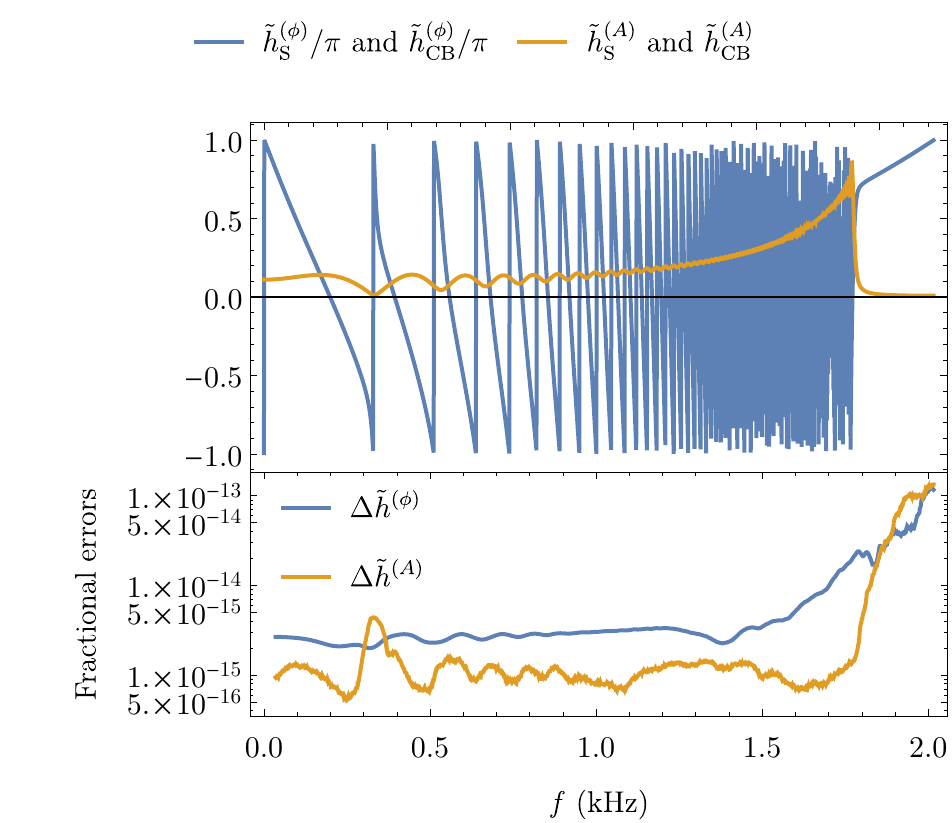}
	\caption{\label{fig:pointwise-error} 
		\emph{Top panel:} The amplitude and the phase part of the waveform associated with $l = 1$. There is visual agreement among the fiducial \emph{CBwaves} waveform and its surrogate prediction throughout the entire frequency range. \emph{Bottom panel:} The relative errors \eqref{relative-error} with moving average of $50$ points, defined by eq. \eqref{relative-error}, in the amplitude and the phase difference between the fiducial waveform and its surrogate model prediction. The differences are smaller than the errors intrinsic to the surrogate model itself, as well as those of state-of-the-art numerical relativity simulations.}
\end{figure}

A discrete version of the normed difference between a fiducial waveform and its surrogate is what we can actually measure: 
\begin{equation} \label{surr-error}
\Delta\tilde{h}^{2}(\lambda) = f_{s} \sum _{k=0}^{N-1}\left|\lvert\tilde{h}_{\text{CB}}(f_{k}; \lambda) - \tilde{h}_{\text{S}}(f_{k}; \lambda)\right|^{2}
\end{equation}
where $f_{s}$ is the sampling frequency discussed in Sec. \ref{sec:frequency-grid}. The square of the normed difference between two waveforms, refered to as the \emph{surrogate error}, is directly related to their overlap \eqref{inner-product}. It is the dominant source of error in the surrogate model that translates directly into errors in the fits of the parameters for building the surrogate. \cite{field} Fig. \ref{fig:surr-error} shows the linear correlation of the surrogate error in eq. \eqref{surr-error} with the time spacing $\Delta f$ in the regularly spaced grids (\ref{time-grid}--\ref{FFT-definition}). The surrogate model gradually converges to the fiducial one at finer time scales (i.e. larger sampling frequencies). Other errors of interest are the pointwise ones (separately for the amplitude and phase). They are encoded in the $l$th surrogate model prediction \eqref{inerpolated-waveform} as
\begin{equation} \label{interpolants}
\begin{array}{lll}
\tilde{h}^{(A)}_{\text{S}}(f_{k}; \lambda_{l}) \hspace{-2pt} & \equiv & \hspace{-2pt} I_{f}[\mathcal{B}^{(A)} \cdot I[\mathcal{M}^{(A)}](\lambda_{l})], \\[10pt]
\tilde{h}^{(\phi)}_{\text{S}}(f_{k}; \lambda_{l}) \hspace{-2pt} & \equiv & \hspace{-2pt} I_{f}[\mathcal{B}^{(\Phi)} \cdot I[\mathcal{M}^{(\Phi)}](\lambda_{l})],
\end{array}
\end{equation}
respectively. The \emph{relative errors} in approximating the amplitude and phase of a fiducial waveform by its surrogate model prediction is then expressed by
\begin{equation} \label{relative-error}
\begin{array}{lll}
\Delta\tilde{h}^{(A)}(f_{k}; \lambda_{l}) \hspace{-2pt} & = & \hspace{-2pt} \left|1 - \tilde{h}^{(A)}_{\text{S}}(f_{k}; \lambda_{l})/\tilde{h}^{(A)}_{\text{CB}}(f_{k}; \lambda_{l})\right|, \\[10pt]
\Delta\tilde{h}^{(\phi)}(f_{k}; \lambda_{l}) \hspace{-2pt} & = & \hspace{-2pt}\left|1 - \tilde{h}^{(\phi)}_{\text{S}}(f_{k}; \lambda_{l})/\tilde{h}^{(\phi)}_{\text{CB}}(f_{k}; \lambda_{l})\right|
\end{array}
\end{equation}
where the amplitude and phase parts of fiducial waveforms, $\tilde{h}^{(A)}_{\text{CB}}(f_{k}; \lambda_{l})$ and $\tilde{h}^{(\phi)}_{\text{CB}}(f_{k}; \lambda_{l})$, respectively, are given by eq. \eqref{FFT-amplitude-and-phase} on $N$ discrete frequency points $f_{k}$. \\

Fig. \ref{fig:pointwise-error} shows a comparison between the surrogate and fiducial model, using the template assigned to $l = 1$. The top panel shows that the fiducial and surrogate waveforms are visually indistinguishable. The bottom panel demonstrates that both amplitude and phase pointwise errors \eqref{relative-error} increase with frequency. Nevertheless, the errors are indeed as small as predicted on Fig. \ref{fig:sigma}. A moving average of $50$ points was used to smooth out short-term fluctuations  in the error and highlight longer-term trends.

\subsection{Computational cost and speedup for surrogate model predictions}
Apart from the requirements for accuracy or reliability, a ROM building is considered efficient if it generates cost-efficient surrogate models. The major advantage of using surrogate model predictions in lieu of actual waveform evaluations is their significantly reduced resource consumption. Now we discuss the computational cost, in terms of \emph{operation counts} and \emph{run-time}, of ROM building and present the desired \emph{speedup} that can be achieved when evaluating surrogate models. \\

As described in Sec. \ref{sec:surrogate}, the complete surrogate model \eqref{inerpolated-waveform} is assembled with the evaluation of $r$ projection coefficients $\mu_{l}(f)$ given in \eqref{projection-coefficient} and $2r$ fitting functions $\{\tilde{h}^{(A)}_{l}(\lambda)\}_{l=1}^{r}$ and $\{\tilde{h}^{(\phi)}_{l}(\lambda)\}_{l=1}^{r}$ given in \eqref{interpolants}. In order to construct a surrogate model for some parameter $\lambda$, one only needs to evaluate each of those $2r$ fitting functions at $\lambda_{0}$, recover the $r$ complex values $\{\tilde{h}_{l}^{(A)}(\lambda_{0})\exp[-i \tilde{h}_{l}^{(\phi)}(\lambda_{0})]\}_{l=1}^{r}$, and perform the summation over the index $l$. Each $\mu_{l}(f)$ is a complex-valued frequency series with $N$ samples. Therefore, the total operation count to evaluate the surrogate model at each $\lambda_{0}$ is $(2r-1)N$ plus the cost to evaluate the fitting functions. \cite{field} The entire process of constructing a small, efficient ROM which is comprized of only $r = 550$ waveform templates sampled at $N = 4000$ grid points requires the execution of approximately $4.4 \times 10^{6}$ operations (excluding the cost of evaluating the fitting functions). \\
\begin{figure}[h]
	\includegraphics[scale=0.60]{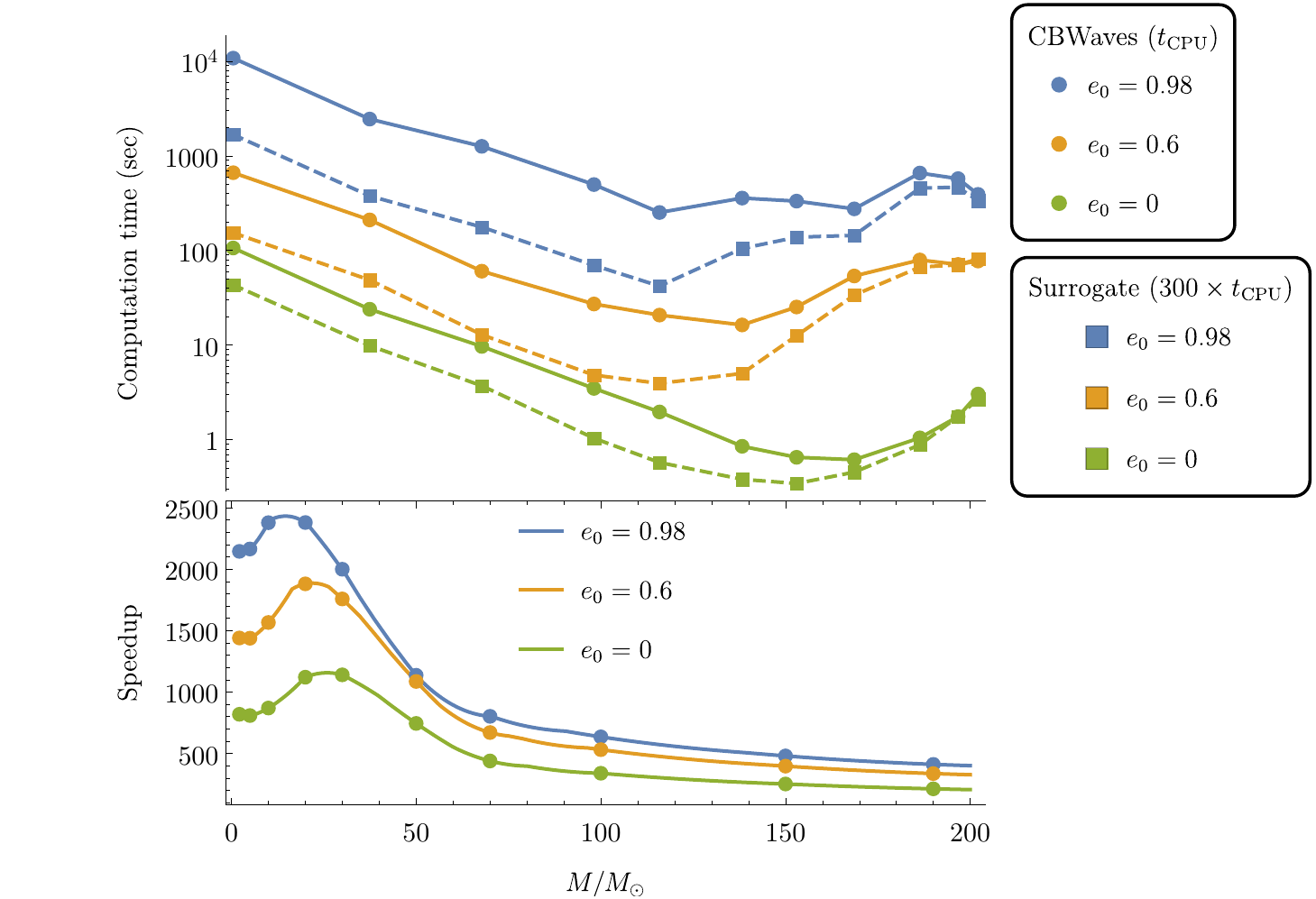}
	\caption{\label{fig:speedup} 
		\emph{Top panel:} Computational time $t_{\text{CPU}}$ to generate fiducial waveforms by CBwaves code (dots; connected by solid lines) against the cost of evaluating corresponding surrogates by ROM (rectangles; connected by dashed lines). The computational time was measured for three different initial eccentricities of equal-mass configurations, each  associated with different colours. \emph{Bottom panel:} The speedup in evaluating the surrogate model is several thousand times faster around $10-50$ $M_{\odot}$ than generating CBwaves waveforms. For high total mass the speedup falls off to several hundreds. The speedup is roughly twice as great for configurations having extremely high initial eccentricity at $e_{0} = 0.98$ (blue line) as for circular ones at $e_{0} = 0$ (green line).}
\end{figure}

The notion of `speedup', in our terminology, is the number that evaluates the relative performance of generating the same waveforms on the same processor by the execution of CBwaves code and of the surrogate model. More specifically, we test the acceleration of waveform generation by measures on the length of time required to perform each computational process. Let as note that the time which was denoted by $t_{\text{int}}$ and was referred to as `integration run-time' in Sec. \ref{sec:eccentric-waveform} is actually the execution time during which the processor is actively working on our computations. It is referred to as CPU time (or run-time) and will be denoted by $t_{\text{CPU}}$. In contrast, the actual elapsed real time accounts for the whole duration from when the computational process was started until the time it terminated. The difference between the two can arise from architecture and run-time dependent factors such as waiting for input/output operations (e.g. saving waveform templates). Consequently, the elapsed real time is greater than or equal to the CPU time.

Fig. \ref{fig:speedup} shows (on top) the computation time or CPU time for CBwaves waveforms (solid lines) against corresponding surrogate waveforms (dashed lines) as a function of total mass of the binary system. The total mass $M$ is measured in the same 11 points as in Fig. \ref{fig:Tgen} for three different initial eccentricities ($e_{0} = \{0.98,\, 0.6,\, 0\}$) of equal-mass configurations, each associated with different colours. The computation time $t_{\text{CPU}}$ for surrogates is multiplied by a factor of $300$ in order to shift the curves close to their respective CBwaves counterparts and enable visual comparison. The bottom panel demonstrates that surrogates are several thousand times faster around $10-50$ $M_{\odot}$ to evaluate as compared to the cost of generating CBwaves waveforms. The speedup falls off to several hundreds as the total mass increases. Moreover, the speedup grows when the initial eccentricity $e_{0}$ is increased in much the same way as with the mass disparity $\bar{q}$ (cf. Fig. 6 in \cite{purrer2}). The speedup is roughly twice as great for configurations having extremely high initial eccentricity ($e_{0} = 0.98$) as for circular ones ($e_{0} = 0$). The resemblance of the influence of $e_{0}$ and $\bar{q}$ exercize on the speedup can be attributed to their asymptotical nature as it had been pointed out earlier in Sec. \ref{sec:eccentric-waveform}. It is also evident that the speedup culminates when waveforms for configurations of very low total mass and very high eccentricity are generated. Such waveforms are prohibitively expensive to generate with CBwaves in contrast to surrogates that are generated at the same cost, regardless of the parameters of the configuration.

Let us note that successive versions of SEOBNR ROMs have been developed and put to use within LAL (LSC Algorithms Library) to shorten data analysis applications carried out since the first observation runs have begun. \cite{veitch} It has been shown in \cite{field,field2} that the cost of evaluating the surrogate model is linear in the number of samples $N$ (cf. our Fig. \ref{fig:surr-error} where the surrogate error depends on the sampling rate). Depending on the sampling rate, the speedup is between 2 and almost 4 orders of magnitude. The speedup in evaluating surrogate models compared to generating NR waveforms with the LAL analysis routines is crucial for searches and theoretical parameter estimations. SEOBNR (aligned-spin Effective-One-Body), IMRPhenomD (Inspiral-Merger-Ringdown Phenomenological Model `D') and PhenSpinTaylorRD waveform approximants are among the best available GW models for generic spinning, compact binaries. In comparison with our results, the speedup achieved at the typical rate of $2.048$ kHz used by aLIGO and aVirgo observatories is roughly $2300$. \cite{LIGO}

\section{Concluding remarks and outlook} \label{sec:conclusion}
The primary goals of the present paper have been to propose a potential extension of the ROM techniques to alleviate the computational burden of constructing waveform templates for coalescing compact binaries with any residual orbital eccentricity and to validate the applicability of ROMs to this particular family of waveforms. ROMs have been applied to several waveform families (SEOBNR, IMRPhenomP and PhenSpinTaylorRD) in LAL routines for gravitational-wave data analysis. \cite{field2,field3,cannon,cannon2,cannon3,cannon4,maday,privitera} The aforementioned waveform families provide efficient descriptions of gravitational waves emitted during the late IMR (inspiral, merger, and ringdown) stages of compact binary systems, but only in the zero-eccentricity limit. The major motivation for extending the scope of application beyond the zero-eccentricity limit is based on the ground, referred to in Sec. \ref{sec:intro}, that the great majority of compact objects formed in dense stellar environments retain some non-negligible eccentricity when entering the frequency band of ground-based GW detectors \cite{peters,antonini}, as well as the impact of eccentricity on the accuracy of parameter estimation for BNSs \cite{favata}.

Our approach to construct frequency-domain ROMs has been predominantly based on the method outlined in Refs. \cite{purrer,purrer2} (see Sec. \ref{sec:rom}). Input waveforms comprised in the ROM are Fourier transformed and split into their amplitude and phase parts (see Sec. \ref{sec:fourier-transform}). These functions are accuretely represented on a common, finely sampled and regularly spaced frequency grid defined in Sec. \ref{sec:frequency-grid} with only $N = 4000$ equidistant nodes, with a sampling frequency recorded well above the required Nyquist frequency, at $f_{s} = 16.384$ kHz. Fig. \ref{fig:surr-error} demonstrates that, beside the degree of model order reduction, the accuracy of surrogate-waveform representation relies on the sampling frequency. The upper and lower limits of frequency contained in the grid are determined from the ISCO frequency for the lowest total-mass configuration of interest (which is roughly $2$ kHz in the present work) and the low-frequency cutoff of the detector noise spectrum (which is close to $10$ Hz for aLIGO design). The ROM is designed to be capable of producing surrogates for GWs from coalescing compact binaries of total mass between $2.15M_{\odot}$ and $215M_{\odot}$, thereby covering the entire total-mass range of stellar-mass BBH/BNS systems of interest for ground-based GW detectors. The mass ratio is allowed to range between equal mass at $q = 1$ and relatively high mass-ratio at $q \approx 0.01$ while the initial orbital eccentricity changes over a relatively wide range of values from $e_{0} = 0$ (circular orbits) up to $e_{0} = 0.95$ (highly eccentric orbits). Configurations with both low total-mass and high mass-ratio would imply component masses well bellow $1M_{\odot}$, which, of course, are excluded as inconceivable astrophysical sources. Despite the fact that the investigation has been restricted to a feasible 3-dimensional subset of the full 8-dimensional parameter space of GW signals (see Fig. \ref{fig:parameter-space}), the conclusions of Sec. \ref{sec:rom}, in agreement with that of Refs. \cite{field,field2,field3,purrer,purrer2}, suggest that a full representation of the 8-parameter space might actually be achievable with a relatively compact reduced basis (cf. Ref. \cite{field}). Template placement algorithms based on template-space metric (such as in Ref. \cite{kalaghatgi,cokelaer}) make admittedly far more effective coverage of the parameter space than the uniform spacings we used in this preliminary study. As a matter of fact, Fig. \ref{fig:parameterspace} illustrates that the large majority of parameters of the selected templates constituting the reduced basis are concentrated along the axes of the parameter space.\\

The reduced bases were built separatelly for the input amplitude and phase (see Fig. \ref{fig:BasisFunc}) by the decomposition of template matrices that comprise 550, 936, and 1800 input waveforms, respectively. The projection coefficients for corresponding input waveforms projected onto their reduced bases, were calculated as functions of the model parameters $(M,\, q,\, e_{0})$ and were interpolated by tensor product cubic lines over the parameter space. Finally, the ROM which preserves fundamental features of the original full-order model is assembled from its constituent parts. Fig. \ref{fig:sigma} demonstrates the underlying hierarchical nature of the generated template banks and indicates that the truncation error in the approximated representation of surrogates decreases with the number of SVD components retained, characterized by a rate at which SVs decrease. Extremely little ($r \lesssim 400$) or large number ($r \gtrsim 500-600$) of SVD components retained are equally poor choices because the amount of information is either insufficient to construct accurate representations or excessively large compared to the achieved accuracy. An effective rank is chosen preferentially from a ROM which posess the lowest SV with the smallest possible number of components retained (in our case $r = 456$). The first part of Sec. \ref{sec:efficiency} assess the error of surrogate model predictions for waveforms that were originally not present in the original template bank, with special regard to the impact of frequency on the reconstruction error. To that end, reference waveforms were generated by CBwaves in all the intersection points right between the grid poinst of the original template bank (see the yellow in Fig. \ref{fig:parameter-space}). Finally, the surrogates were evaluated in the corresponding parameter-space points for comparison and the relative error was measured along all the $N = 4000$ frequency points. The bottom panel of Fig. \ref{fig:pointwise-error} attests that the relative error of the approximated representation is consistent with the error estimates derived from the singular values $(\Delta\tilde{h}^{(A)} \approx 10^{-15},\, \Delta\tilde{h}^{(\phi)} \approx  10^{-13})$ over a large portion of the frequency range, but larger than expected at around the starting frequency $(\Delta\tilde{h}^{(A)} \approx 10^{-13},\, \Delta\tilde{h}^{(\phi)} \approx  10^{-13})$. The figure indicates that the relative error of the amplitude and phase increases with the frequency. Our results provide clear examples of the construction and use of ROMs for eccentric inspiral waveforms. 

Our results also provide strong evidence that large increases in the speed of computation are obtained through the use of ROMs. Fig. \ref{fig:Tgen} has exposed that the cost of computating input waveforms increases exponentially as the total mass decreases, but rises asymptotically at an even faster rate than the initial eccentricity or mass disparity increase. In contrast to the cost of EOB waveform (full IMR) generation that rises steeply as the starting frequency is decreased (see Ref. \cite{purrer2}), the cost of CBwaves waveform (inspiral-only) generation rises more gradually. The cost of input waveform generation varies considerably in the region of parameter space $(M,\, q,\, e_{0})$ explored and depicted in Fig. \ref{fig:parameter-space}, but Fig. \ref{fig:freqdist} has revealed that only a surprisingly small fraction of waveforms of high-eccentricity and high-mass-disparity configurations are actually responsible for the prohibitively large time-consumption of integrating a large number of 3PN-accurate equation of motion over the investigated range of parameters. As discussed in the second part of Sec. \ref{sec:efficiency} (based on Ref. \cite{purrer}), the cost of generating surrogate waveforms (shown in the top panel of Fig. \ref{fig:speedup}) comprises a constant cost of the spline interpolation at each frequency point and a cost of performing the interpolations of coefficients over the parameter space. The speedup in evaluating the surrogate model, shown in the bottom panel of Fig. \ref{fig:speedup}, is 2--3 orders of magnitude faster than generating corresponding CBwaves waveforms overall, reaching a factor of several thousand around 10--50 $M_{\odot}$. 

Finally, the method presented in this paper is limited to building surrogate models of inspiral-only PN input waveforms for the reason that eccentric binaries circularize in the last few cycles before the merger. Nevertheless, composite waveforms that fully cover all the IMR stages can be constructed as prescribed in Ref. \cite{purrer,purrer2} by matching the inspiral and NR waveforms of merger stages in either the time or frequency domain and then fitting this `hybrid' waveform to the ring-down part, described by damped exponentials. The gap between the initial part of the waveform and its final ring-down part, described by damped exponentials, is bridged by a phenomenological phase. The practical implementations of `hybrid' waveforms that comprise eccentric inspirals of will be left for future work. We anticipate substantial speedup factors to come for predicting NR waveforms with a surrogate model compared to the expensive numerical simulations for the same parameters. Developing an efficient template placement technique (such as in Ref. \cite{kalaghatgi,cokelaer}) for better coverage of the parameter space and an adaptive sampling technique in the frequency domain are critical factors in the operational efficiency of ROMs and have been left for future work. All these ultimately leading to computationally feasible and successful exploration of the full 8-dimensional parameter space of GW signals.

\begin{acknowledgments}
This work was supported by the NKFIH 124366 and NKFIH 124508 grants. Partial support comes from NewCompStar, COST Action Program MP1304.
\end{acknowledgments}

\bibliography{cikk_rb}

\end{document}